\documentclass[12pt,preprint]{aastex}
\usepackage{float,epsfig,psfig}

\newcommand{\LA}{\mbox{\raisebox{-0.6ex}{$\stackrel{\textstyle<}{\sim}$}}}
\newcommand{\GA}{\mbox{\raisebox{-0.6ex}{$\stackrel{\textstyle>}{\sim}$}}}
\newcommand{\cxo}{{\sl Chandra}}

\newcommand{\cha}{{\sl Chandra}}

\newcommand{\hst}{{\sl Hubble}}
\newcommand{\msun}{M$_{\odot}$}
\newcommand{\ergl}{ergs~s$^{-1}$}

\newcommand{\fxfo}{$F_{\rm X}/F_{\rm O}$}

\newcommand{\lb}{$L_{\rm B}$}
\newcommand{\etal}{et al.}

\begin{document}

\title{
A Multi-wavelength Study of the 
X-ray Sources in NGC~5018}

\author{
Kajal~K.~Ghosh\altaffilmark{1},
Douglas~A.~Swartz\altaffilmark{1},
Allyn~F.~Tennant\altaffilmark{2},
Kinwah~Wu\altaffilmark{3}, and
Lakshmi~Saripalli\altaffilmark{4}
}
\altaffiltext{1}{Universities Space Research Association,
NASA Marshall Space Flight Center, SD50, Huntsville, AL, USA}
\altaffiltext{2}{Space Science Department,
NASA Marshall Space Flight Center, SD50, Huntsville, AL, USA}
\altaffiltext{3}{MSSL, University College London, Holmbury St. Mary, Surrey,
RH5 6NT, UK}
\altaffiltext{4}{Australia Telescope National Facility, CSIRO, Locked Bag 194, Narrabri, NSW 2390, Australia}

\begin{abstract}
The E3 giant elliptical galaxy NGC~5018 was observed with
 the \cxo\ X-ray Observatory's Advanced CCD Imaging Spectrometer for
 30~ks on 14 April 2001. 
Results of analysis of these X-ray data as well as of complementary optical,
 infrared, and radio data are reported. 
Seven X-ray point sources, including the nucleus, were detected. 
If they are intrinsic to NGC~5018, then all six non-nuclear sources have luminosities
 exceeding $10^{39}$~\ergl\ in the 0.5--8.0~keV energy band; placing them in 
 the class of Ultra-luminous X-ray sources. 
Comparison of X-ray source positions to archival 
 Hubble Space Telescope/Wide Field Planetary Camera 2 (\hst/WFPC2) 
 images reveal four of the six non-nuclear
 sources are spatially--coincident with bright, M$_{V}$\LA -8.6 mag, objects.
These four objects have optical magnitudes and (V--I) colors consistent with
 globular clusters in NGC~5018 but they also have X-ray-to-optical flux ratios 
 consistent with background active galactic nuclei.
Strong, unpolarized, radio emission has been detected from one 
 of the optically-bright counterparts.
Another optically-bright counterpart was observed to vary by $\sim$1 mag in 
 optical observations taken 28 July 1997 and 04 Feb 1999.
Extended X-ray emission is detected within a $\sim$15\arcsec\ radius of the 
 galaxy center at a luminosity of $\sim$$10^{40}$~\ergl\ in the X-ray band. 
Its thermal X-ray spectrum ($kT$$\sim$0.4~keV) and its spatial 
 coincidence with strong H$\alpha$ emission are consistent with a hot gas 
 origin. 
The nucleus itself may be a weak X-ray source,  
 L$_{\rm X}~\LA~3.5$$\times$10$^{39}$~\ergl,
 that displays a radio spectrum typical of AGN.

\end{abstract}

\section{Introduction}

\cxo\ images of nearby galaxies routinely reveal dozens of discrete X-ray
 sources per galaxy above a detection threshold of
 $\sim$$10^{37}$~\ergl\ (Fabbiano \& White 2003).
For many of these images, the population of Ultra-Luminous X-ray sources is
 a major focus of study
 (e.g., Roberts, \& Warwick 2000; Kilgard \etal\ 2002; Humphrey \etal\ 2003;
 Irwin, Athey, \& Bregman 2003; Irwin, Bregman, \& Athey  2004;
 Colbert \etal\ 2004; Swartz \etal\ 2004).

Ultra-Luminous X-ray sources (ULXs) in nearby galaxies
 are non-nuclear point-like objects
 with X-ray luminosities $>$$10^{39}$~\ergl, above the Eddington
 limit for spherical accretion of hydrogen onto a \LA 8~\msun\ object.
ULXs must either accrete mass at a high
 rate or result from supernovae in very favorable circumstances
 in order to emit so copiously in the X-ray band.
If ULXs are accretion-powered,
 then they are either massive~--~the elusive intermediate-mass
 black holes (IMBHs)(Colbert, \& Mushotzky 1999; Makishima 
\etal\ 2000; Colbert \& Ptak 2002; van der Marel 2003), or they emit 
anisotropically (with a true luminosity less than the isotropic equivalent) 
(King \etal\ 2001; 
Georganopoulos, Aharonian, \& Kirk 2002) 
 or they are truly super-Eddington (Abramowicz \etal\ 1988; Begelman 2002; 
Grimm, Gilfanov, \& Sunyaev 2003). 
Alternatively, some ULXs, at least, may not be accreting objects but 
are associated with supernovae and hypernovae (Franco \etal\ 1993; Plewa 1995).

There is a growing consensus
 (Grimm, Gilfanov, \& Sunyaev 2003; Colbert \etal\ 2004; Fabbiano 2004; Gilfanov 2004; Irwin, Bregman, \& Athey  2004;
 Swartz \etal\ 2004),
 based on global properties of ULX host galaxies,
 that there are (at least) two populations of ULX:
(1) A bright population associated with recent star formation with
 the number of ULXs and their X-ray luminosities scaling with their host
 galaxy's far-infrared luminosity and other tracers of star-formation rate;
and (2) A population of weaker ULXs (perhaps not exceeding
 $\sim$$2\times 10^{39}$ \ergl, Irwin, Bregman, \& Athey  2004)
 found in early-type galaxies with a population proportional to blue
 luminosity and other galaxy mass indicators.
Thus, ULXs are likely to be a heterogeneous group of high-mass X-ray binaries
 (XRBs) and supernovae, products of recent star formation and found
 primarily in late-type galaxies,
 and low-mass XRBs (LMXBs) in the late evolutionary stages of the mass donor
 star, accounting for ULXs in ellipticals (see also King 2002).

This dichotomy implies the ULX population in the E3 galaxy NGC~5018 should
 be composed of LMXBs
 of luminosity \LA $2\times 10^{39}$ \ergl\ with the
 number of ULXs scaling with blue luminosity, \lb.
We find twice as many ULX candidates as expected and
 3 of the 6 candidates have observed luminosities $>$$2\times 10^{39}$~\ergl\ 
in the 0.5-8.0 keV band.
While not beyond statistical chance, this result was remarkable enough
 that closer inspection of the ULX population of NGC~5018 was undertaken
 and is reported here.
Archival \cxo, \hst, 2MASS, newly acquired radio observations 
and data reduction techniques are
 presented in \S~2.
Individual ULX candidates are described
 in \S~3 and the nature of the nucleus and diffuse emission is 
 considered in \S~4.
The possible origin of the X-ray emission and its implication for the recent evolution of NGC~5018 is discussed in \S~5. 

\section{NGC~5018 Observations and Data Reduction}

The type E3 giant elliptical galaxy
 NGC~5018 is the brightest member ($M_B$$=$$-21.45$ mag) of a group 
 of 5 galaxies
 located at a distance of 40.8 Mpc (1\arcsec$=$198~pc).
NGC~5018 is connected to its
 two nearest neighbors, NGC~5022 and MCG~03-34-013 (see Figure~1), by a large 
 H~I bridge
 that supplies gas to NGC~5018 (Kim \etal\ 1988; Guhathakurta \etal\ 1989).
There is also a distinct stellar bridge 
 between NGC~5018 and NGC~5022 and an embedded
 dust lane in NGC~5018; both relics of a past interaction 
 (Malin \& Hadley 1997).
Further evidence of its dynamic history are shells around the galaxy,
 a tidal tail (Malin \& Carter 1983), and a compact radio core 
 (Mollenhoff \etal\ 1992).
The interaction is estimated to have occurred 300 to 600 Myr ago
 (Guhathakurta et al. 1989). 

\clearpage
\begin{center}
\includegraphics[angle=-90,width=\columnwidth]{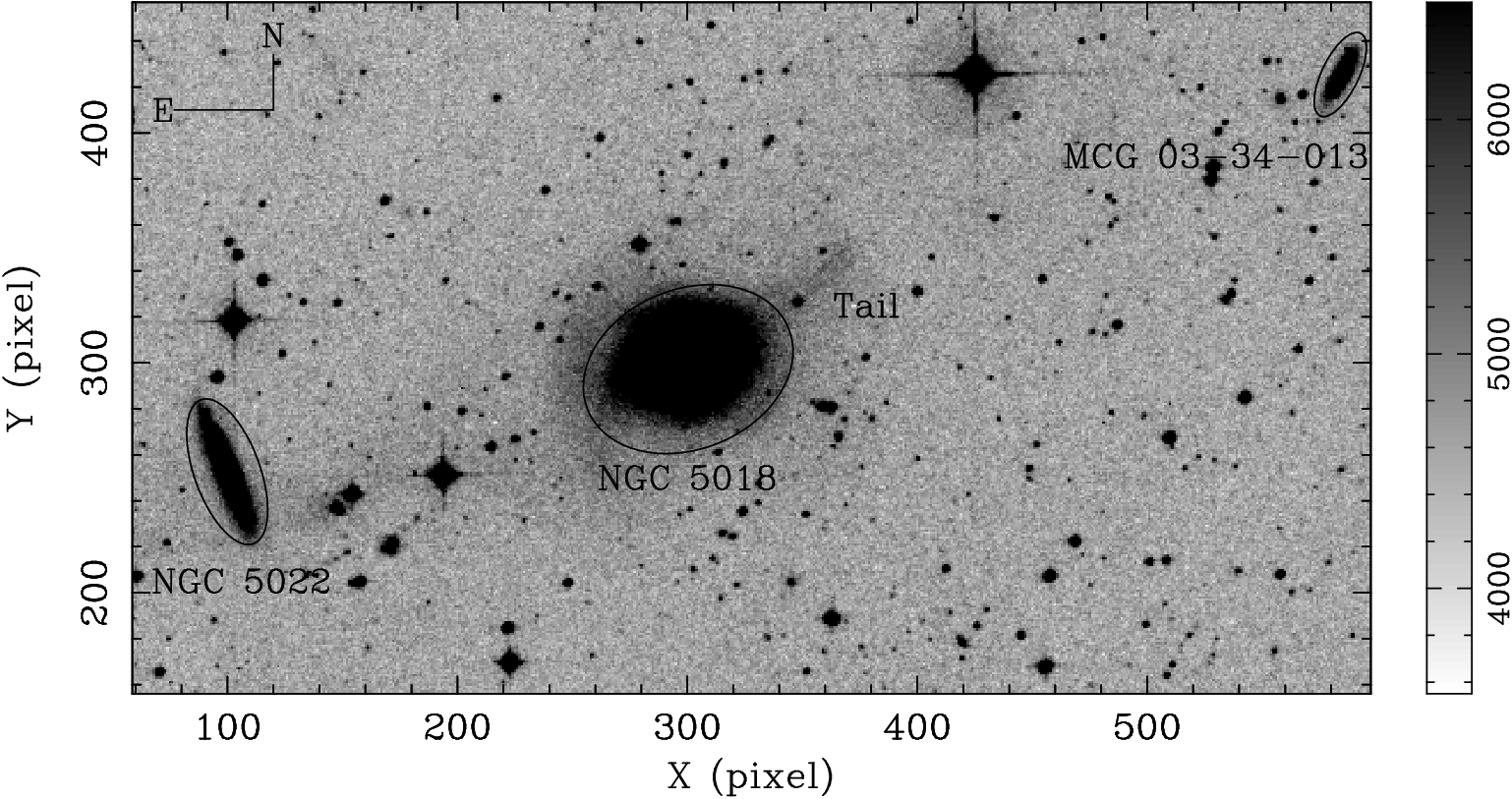}
\figcaption{Optical Digitized Sky Survey (red) image of NGC~5018. A stellar bridge lies between 
NGC~5018 and NGC~5022 and a tidal tail is present on the opposite side of this bridge.
D$_{25}$ isophotes (defined as the major isophotal angular diameter measured at a
blue light surface brightness level of 25.0 mag sec$^{-2}$) are shown around NGC~5022, NGC~5018 and MCG 03-34-013. The image is approximately 16.4 x 10.5 arcmin}
\end{center}
\clearpage
\subsection{X-ray Observations}

A 30 kilosecond observation of NGC~5018 was carried out with the \cxo\
 Advanced CCD Imaging Spectrometer (ACIS) operating in imaging mode on
 2001 April 14 (ObsID 2070).
This dataset was retrieved from the \cxo\ archive
 and the Level~2 event list was used to extract the events within the
 3.\arcmin3 $D_{25}$ isophote of NGC~5018 
(fully within the back-illuminated CCD S3).
Source detections and source and background light curves and spectra were
 extracted using the locally-developed software package {\tt lextrct}
 (Tennant \etal\ 2005).
Some details of data reduction methods are given in Swartz et al. (2003).

The detection limit for point sources in this observation
 is $\sim$$6\times 10^{38}$ \ergl, assuming an
 absorbed power-law of photon index $\Gamma=1.8$
 (the average for ULXs in the sample considered by Swartz \etal\ 2004),
 the Galactic absorption column along the line of sight
 ($N_H$$=$$6.98\times 10^{20}$~cm$^{-2}$),
 and a 10~c detection limit.
Figure 2 shows the \cxo\ image of NGC~5018 in the 0.5 to 8.0~keV band
 with 2\arcsec\ radius circles about the detected source locations.
Seven sources (including the nucleus) were detected within the $D_{25}$ region
above a signal--to--noise ratio of 2.8. 
Table~1 lists the \cxo\ positions, X-ray spectral fit parameters,
 0.5--8.0~keV absorption-corrected luminosities, X-ray colors,
 and total detected counts
 for each source.
 
\clearpage
\begin{center}
\includegraphics[angle=-90,width=\columnwidth]{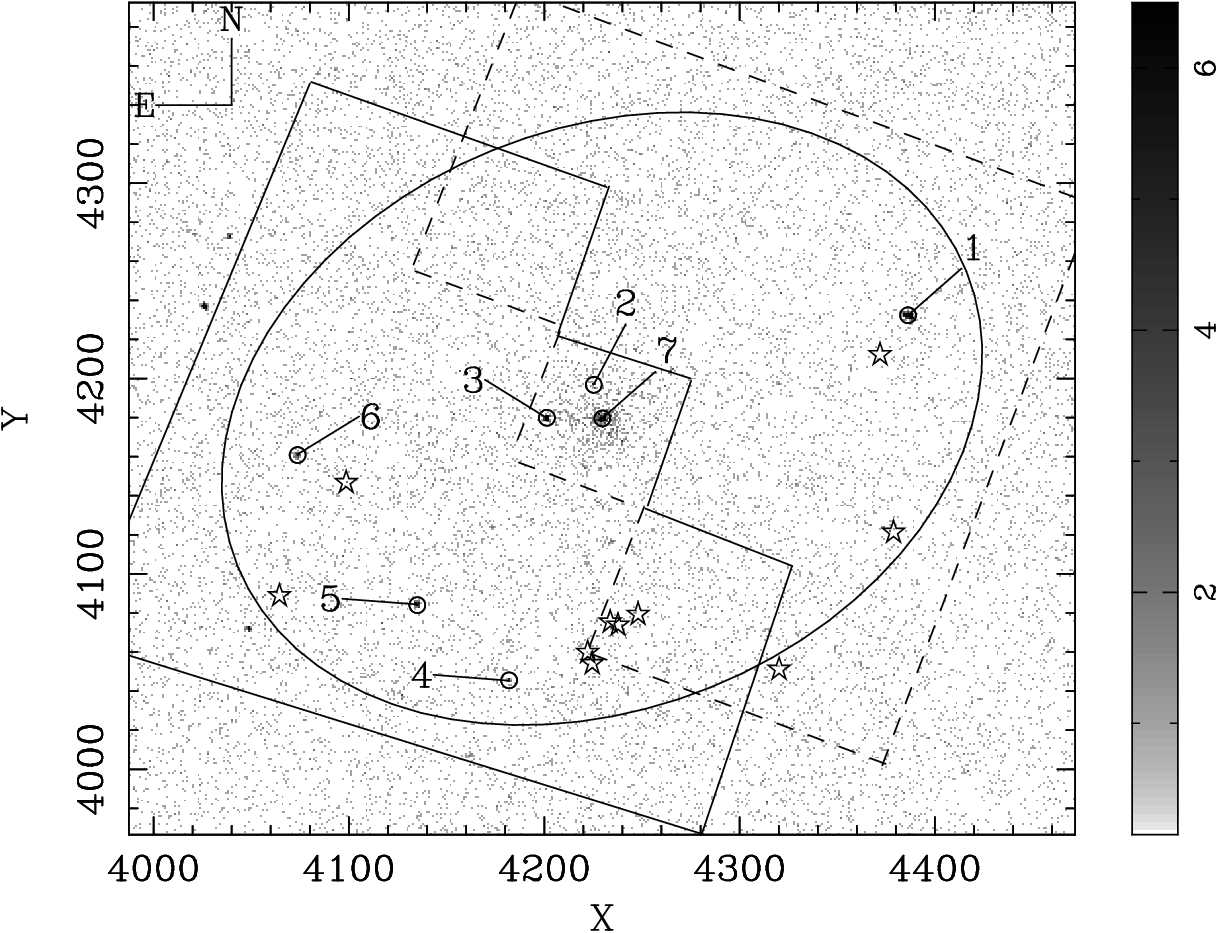}
\figcaption{\cxo\ image of NGC~5018 in the 0.5 to 8.0~keV band.
The ellipse denotes the D$_{25}$$=$3.3\arcmin\ isophote.
The bright nucleus is surrounded by diffuse emission extending to 
$\sim$10\arcsec.
The seven detected point sources are indicated with 2\arcsec\ radius circles. 
Solid lines depict the boundaries of available \hst/WFPC2 images observed with F555W (4400 s) and F814W (5240 s) filters and dashed lines depict the orientation used for 
F336W (1800 s), F555W (1200 s), and F814W (1900 s) images, see Table 2. 
Ten USNO astrometric stars are shown with star symbols. (Two other USNO stars lying
very close to the nucleus are not shown) The image is approximately 4.1 x 3.5 arcmin.
}
\end{center}
\clearpage

The centroids of the source positions were determined by fitting an
elliptical Gaussian to the spatial distribution of X-ray events.
The estimated statistical uncertainty in the source positions was computed from the
ratio of the size of the point spread function and the square root of the 
source counts. These values are between 0.\arcsec05 and 0.\arcsec13.

The 3 brightest sources, including the nucleus,
 produced enough X-ray counts for spectral analysis.
For these sources, spectral redistribution matrices and ancillary response
 files were generated using CIAO, v2.3, and models were fitted to
 their 0.5 to 8.0~keV energy spectra, binned to obtain at least 10~c
 per fitting bin, using XSPEC v11.3.
Parameters for the best-fitting models are listed in Table~1 and discussed
 more fully in \S~3 and~4. 
The X-ray light curves for these three sources were also extracted,
 using 500~s binning, and tested against the constant countrate hypothesis.

The 0.5-8.0~keV X-ray luminosities are estimated from model
 fits to the brightest sources or by assuming a $\Gamma=1.8$ absorbed
 power law and the Galactic N$_{H}$ value of 7$\times$10$^{20}$~cm$^{-2}$ in the direction of NGC~5018, for the weaker sources using PIMMS (Mukai 1993).  

\subsection{Optical Observations}

A log of archival \hst/WFPC2 observations of NGC~5018 is given in Table~2. 
Exposure calibrated and co-added cosmic ray free WFPC2 images were retrieved from the WFPC2-Associations\footnote
{http://archive.stsci.edu/hst/wfpc2/} database for analysis.
Positions of the \hst\ images of NGC~5018 relative to the \cxo\ image are 
 outlined in Figure~2.

The nucleus is the only catalogued object common between the \hst\ and \cha\ images
 and it was used to register the two images assuming the centroids of the 
 optical and X-ray emission coincide. The uncertainties of the best-fit elliptical Gaussian to the X-ray nucleus is $\sim$0.\arcsec 05 and we take this to be the 
 registration uncertainty for the two ULX candidates lying on the 
 \hst/PC1 image.
This uncertainty is combined in quadrature with the (statistical) uncertainties
 in the X-ray positions of these two sources 
 to give their final positional uncertainties as listed in Table~3.
 
For the remaining ULX candidates there is an additional uncertainty in the 
 registration because these sources lie on the outer \hst/WF (WF2, WF3, 
 and WF4) CCDs.
This additional uncertainty was estimated using the METRIC program\footnote{http://stsdas.stsci.edu/cgi-bin/gethelp.cgi?metric.hlp}.
The images of the ten USNO stars on these CCDs along with two USNO stars on PC1
 were used to determine this uncertainty.
First we noted the differences between the USNO catalog and the \hst\ values of the 
 positions of the two USNO stars on the PC1 image and computed the average offset.
  We then applied this average value as the offset to the WF images
and determined the differences between the catalog and the \hst\ values of the USNO 
stars located on the WF images. These values are then considered as the relative 
positional errors between the PC and WF images, which range between 0.\arcsec 17 
and 0.\arcsec 33 for the three CCDs. This results in slightly larger uncertainties
for the ULX candidates lying on these CCDs as can be seen from Table~3.

A potential optical counterpart was discovered within the resulting error
 circles for four of the candidate ULXs in at least two \hst\ image bands.
The other two ULX candidates have only upper limits in V and I. The upper limit values were determined by centering on the brightest
pixel within or nearest to the error-circle. The difference in counts obtained from a circle of 3-pixel radius and that obtained from an annulus of 3 and 5 pixel radii, were
used as the upper limit count of the optical counterpart of the ULX candidate
The resulting observed magnitudes, aperture-corrected using the
 curve of growth method, are listed in Table~3.
Also listed are the absolute V--band magnitudes assuming distance modulus = 33.1 mag and corrected for Galactic extinction,
 optical colors, and the X-ray--to--optical flux ratios ($F_X/F_O$).
F555W--band magnitudes of the 
ULX candidates were used to compute the values of $F_O$ using the current values 
of PHOTFLAM and Zeropoint in the VEGAMAG system given in the 
\hst\ data handbook for WFPC2\footnote{http://www.stsci.edu/instruments/wfpc2/Wfpc2$\_$dhb/wfpc2$\_$ch52.html~\#~1902177}.
For brevity, we will refer to F336W, F555W, and F814W 
as U, V and I, respectively.

Measured fluxes were consistently lower in the association dataset U3M72505B
when compared to the U3M72508B data.
We downloaded the component datasets that went into both associations
and found consistency at that level.
Thus it appears that the U3M72505B association dataset has been incorrectly
processed and we have ignored this dataset and used our own processing of
its component files.
We also found that the F336W band association data (U2ST0201B) still
showed obvious cosmic ray tracks.
Again we downloaded the component files and reprocessed the data which
effectively removed the cosmic rays.
This did not change the measured flux for ULX 3, but it did produce
a better looking figure.

\subsection{Radio observations }

We have carried out radio observations of NGC~5018 using the Australia Telescope Compact Array (ATCA). 
We measured radio continuum flux data in total and polarized intensity on June 30,  
 2004, simultaneously observing at four frequencies, 1.4, 2.3, 4.8 and 8.6 GHz. 
The nucleus and the ULX candidate \#~6 were both detected at high significance.

In addition,
 low resolution 20 cm continuum and H~I~$\lambda$-21 cm (Kim \etal\ 1988) 
 and high resolution 20 cm continuum observations (Mollenhoff \etal\ 1992) 
 of NGC~5018 were carried out with the Very Large Array of the 
 National Radio Astronomy Observatory. 
These observations also detected radio emission from the nucleus 
 and at the location of source \#~6.
Results of these measurements are given in Table 5.
There are no radio sources in the NRAO/VLA Sky Survey catalog
 near the locations of the other ULX candidates.
These regions have not been covered by the FIRST survey. 
 
\subsection{Infrared observations }

J, H and K$_{s}$ band images of NGC~5018 were retrieved from the 2MASS archive. 
After inspecting these images around the positions of the ULX candidates, 
 we decided to analyze only the K$_{s}$ band image as it is of the highest quality.
The astrometry between the 2MASS and \cxo\ images was done using the
 nucleus as a registration object. 
No prominent sources were detected at the positions of the ULX candidates.

\section{Non-nuclear X-ray sources}

All six non--nuclear X-ray sources are potential ULXs assuming they are 
 within the host galaxy. 
Four sources are spatially--coincident with potential optical counterparts and
 upper limits to the optical brightness are made for the remaining two sources
 as described in the following subsections.
The high measured values of F$_{X}$/F$_{O}$ suggests neither of these two 
 optically-weak sources are
 foreground stars nor background AGNs (Maccacaro \etal\ 1988).
For the four sources with well-defined optical counterparts, 
 F$_{X}$/F$_{O}$$\sim$1 which is typical of AGNs 
 (Green \etal\ 2004; Veron-Cetty \& Veron 2003).
Statistically, we expect about 1.4 background AGNs among our detected sources
 based on the \cxo\ deep field studies (Brandt \etal\ 2001).
However,  as discussed below, the counterpart optical colors and brightnesses 
 are also consistent with globular clusters within NGC~5018.
Before discussing the individual sources in detail, we present X-ray color-color
(\S~3.1) and optical color-magnitude (\S 3.2) diagrams for the entire population
of detected X-ray sources.
 
\subsection{X-ray color-color diagram}

The background-subtracted
 X-ray colors are defined as $(M-S)/T$ and $(H-M)/T$, 
 where $T=S+M+H$ and
 $S$, $M$, and $H$ denote counts in the 0.5-1.0~keV, 1.0-2.0 keV,
 and 2.0-8.0 keV bands, respectively, extracted from $\sim$90\% encircled energy radii. X-ray colors are listed in Table~1. Total detected counts, $T$, are 
given at the last column of Table~1.
Figure 3 displays the X-ray color--color distribution of the seven sources. 
Five sources are located at the position of mildly absorbed 
 X-ray binaries (Prestwich \etal\ 2003) and are otherwise non-descript. 
Source \#~6 appears to be extremely hard and occupies a region in the 
 color-color diagram where the spectrum is very flat (compare Figure~4 
of Prestwich \etal\ 2003). 
This source is also a strong radio source.
The nucleus of NGC~5018, in contrast, is very soft. This is mainly due to the 
presence of a strong thermal component, as discussed in \S~4.

\clearpage
\begin{center}
\includegraphics[angle=-90,width=\columnwidth]{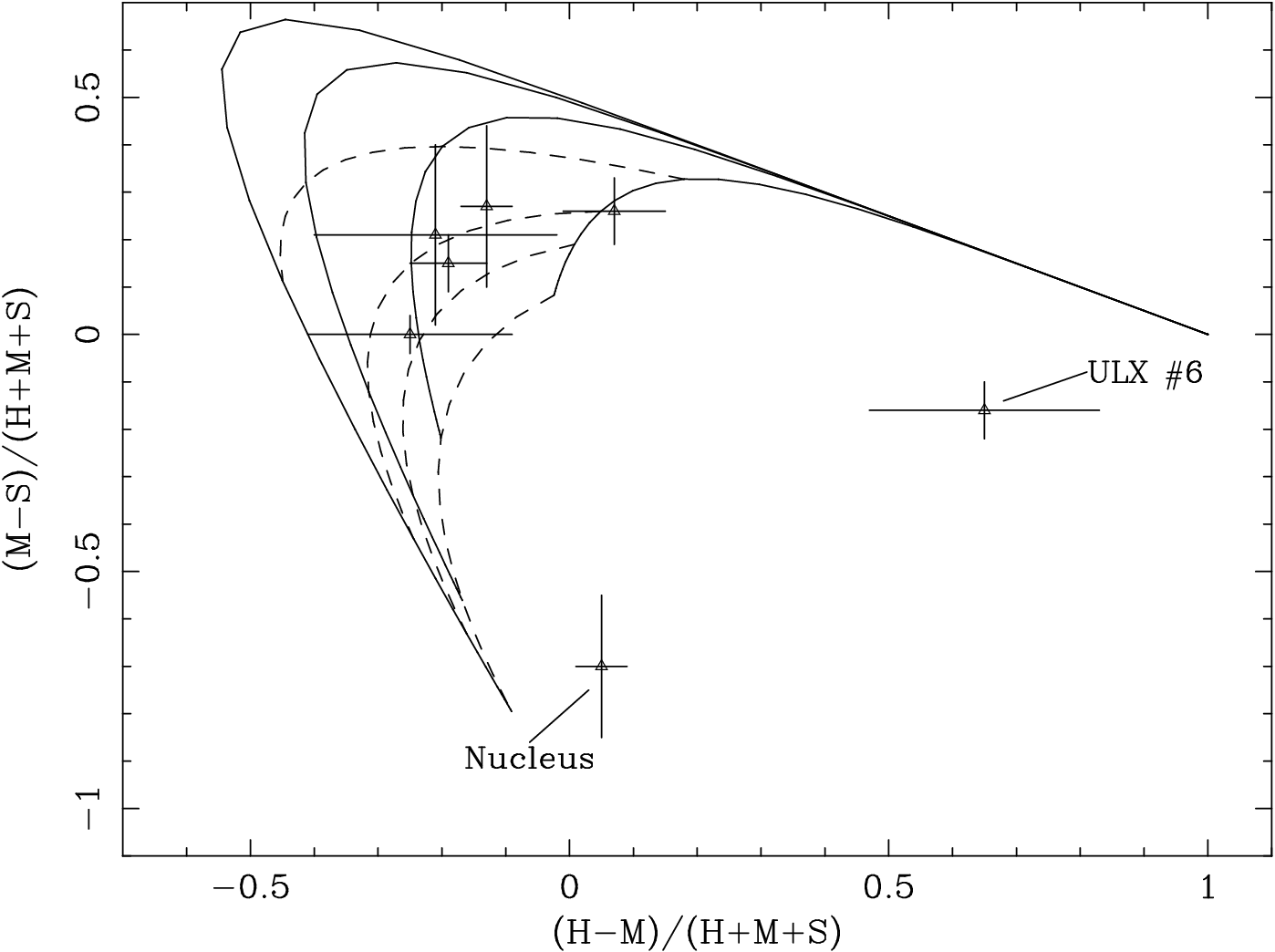}
\figcaption{X-ray color-color diagram of point 
sources in NGC~5018. Solid curves denote colors of absorbed power law models of spectral 
indices $\Gamma$=1,2,3, and 4 (from right to left) and for the range of 
absorbing columns n$_{H}$=10$^{20}$ to 10$^{24}$ cm$^{-2}$. Dashed curves 
denote constant absorption columns of n$_{H}$=10$^{20}$, 10$^{21}$, 2$\times$ 
10$^{21}$, and 5$\times$ 10$^{21}$ cm$^{-2}$ (from bottom to top). Errors shown 
were propagated from the statistical uncertainties in the three X-ray bands.
  }
\end{center}
\clearpage

\subsection{Optical color-magnitude diagram}

The absolute V band magnitude (M$_{V}$) is shown against the (V--I) color 
 for the six possible optical counterparts to the non-nuclear sources in 
 Figure~4. 
Four of the counterparts are much brighter than individual stars. 
All four of these counterparts are as bright or brighter than globular clusters 
 (GCs) in the Milky Way (Harris 1996) indicating they are massive compact
 GCs analogous to some of those found in other nearby galaxies.
In contrast, only upper limits were obtained for any counterparts to
 sources \#~1 and \#~4.

\clearpage
\begin{center}
\includegraphics[angle=-90,width=\columnwidth]{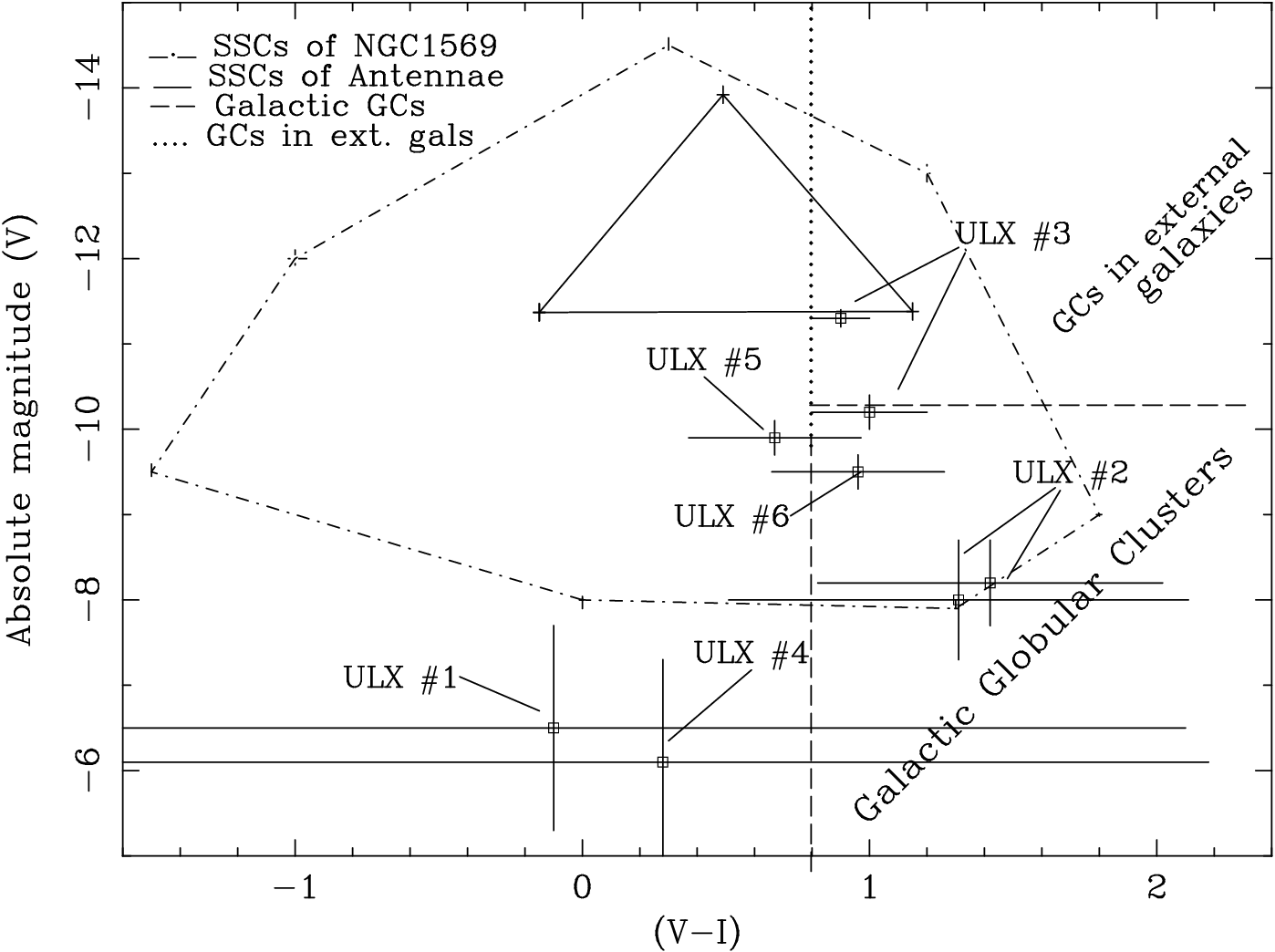}
\figcaption{Absolute magnitude in V--band versus (V--I) color for the 
counterparts of the six non-nuclear X-ray sources detected in NGC~5018. 
Known globular clusters lie to the right and super star clusters 
 to the left, respectively, of a line roughly at (V--I)$=$0.8.
The dashed lines denote the region occupied by Galactic globular clusters
 on the lower right (Harris 1996). 
Globular clusters in other galaxies can be much brighter
 (e.g., Miller \etal\ 1997; Lotz \etal\ 2004) and extend the globular cluster
 region to the upper right. 
Domains of super star clusters detected
in the Antennae (Whitmore \etal\ 1999), and in NGC~1569 (Hunter \etal\ 2000) are also shown. Note the overlap into the red color space occupied by globular clusters.
Two points are plotted for the counterparts to source \#~2 and \#~3 corresponding to
the two seperate HST observations.
}
\end{center}
\clearpage

Also shown in Figure~4 are the regions of color-magnitude space occupied by 
 super star clusters (SSCs) in some well-studied star-forming galaxies.
SSCs are compact, luminous star clusters that have sizes and luminosities
 comparable to GCs but are relatively much younger (\LA 1~Gyr) 
 and bluer than GCs.
They tend to lie blueward of (V--I)$\sim$0.8, as indicated in Figure~4, 
 though there is substantial overlap among the two classes. 
The NGC~5018 GC survey undertaken by Hilker \& Kissler-Patig (1996)
 discovered at least two populations of GC: a blue population with 
 $\langle$V--I$\rangle$$\sim$0.7 (about 10\% of all GCs) and a broader
 red population with (V--I)\GA 1.0.
Hilker \& Kissler-Patig (1996) attribute the blue population to those GCs
 formed within the last few 100~Myr to 6~Gyr
 while the red population is an older primordial population. 
It should be noted that the (V--I) colors alone cannot discriminate between 
 age and metallicity effects. 
The blue clusters may simply be old metal-poor clusters (perhaps captured from
 another galaxy in a merger or interaction).

Four bright optical counterparts to X-ray sources 
 have (V--I) colors overlapping with this population of blue GCs
 in NGC~5018.
They are also within the (V--I) color range of AGNs ($\sim$0~to $\sim$2;
 e.g., Zaggia \etal\ 1998).
In fact, the optical counterpart to source \#~3 (\S~3.5) varied by $\sim$1 mag at
 optical wavelengths suggesting it is a background source and not a GC.

\subsection{ULX Candidate 1}

Source \#~1 is the brightest X-ray object among the six ULX candidates detected in NGC~5018. 
The X-ray spectrum of this source is best described as an absorbed power law. 
Its intrinsic (absorption-corrected) luminosity 
in the 0.5--8.0 keV band is (1.29$\pm$0.17)$\times$10$^{40}$~erg~s$^{-1}$. 
It is not a highly absorbed source, because the value of N$_{H}$ is consistent with the
Galactic N$_{H}$. 

The X-ray lightcurve of source \#~1 is shown in Figure~5.
The source is variable at a 94\% confidence level during the \cxo\ observation
 according to the $\chi^{2}$ test against a constant flux hypothesis
 ($\chi^{2}_r$=1.3 for 62 degrees of freedom).
However, the Kolomogorov-Smirnov test, which is sensitive to longer term
variabiltiy, gave a probability 0.85 indicating the source was almost
constant.
Thus, any short term variability was averaged out over the duration
of the observation.

\clearpage
\begin{center}
\includegraphics[angle=-90,width=\columnwidth]{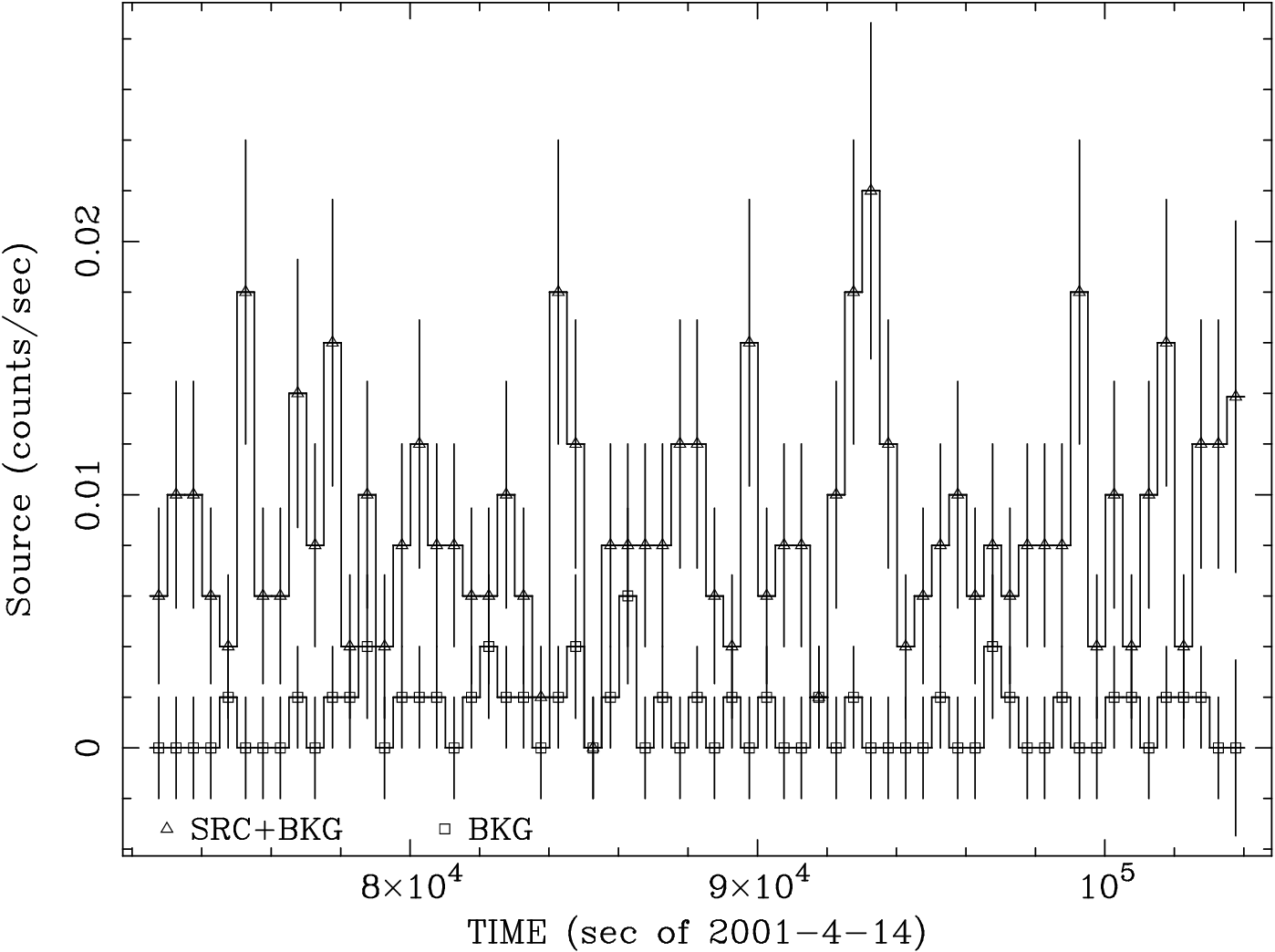}
\figcaption{X-ray lightcurve of source \#~1 with 500~s binning. The upper curve
shows the combined source and background count rate (triangles) and the lower curve
displays the local background count rate.}
\end{center}

\begin{center}
\includegraphics[angle=-90,width=\columnwidth]{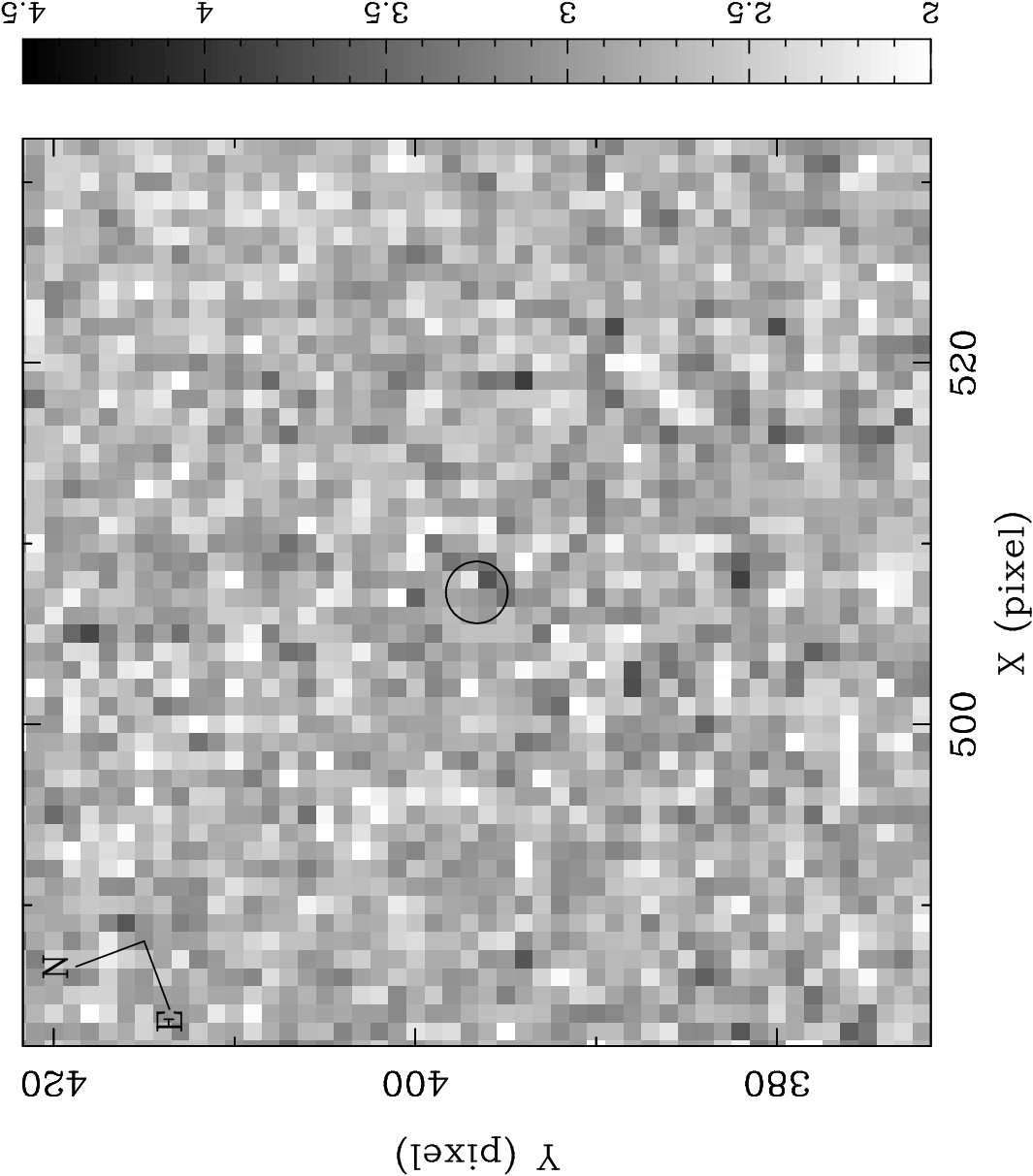}
\figcaption{
\hst\ 5\arcsec$\times$5\arcsec, 1200~s exposure, 
F555W band image centered on the ULX candidate \#~1. 
The 0.17\arcsec\ \cxo\ error circle is also shown. 
There is about a 5 percent chance of finding a source as bright as
the one in the error circle.}
\end{center}
\clearpage

There is a weak source within the X-ray error circle of source \#~1 in both 
 the  V and I \hst\ images (Figure~6).
However, several other nearby pixels show similar brightness levels and the object 
 within the error circle may simply be caused by statistical fluctuations 
 in the NGC~5018 field. It should therefore be considered as an upper limit to 
 any optical counterpart to source \#~1.
Note that the measured absolute magnitude, $M_V$=-6.5$\pm$1.2, still exceeds that
 for all individual stars except for the brightest early-type supergiants.
We note that the faintest globular clusters seen in our Galaxy have a similar brightness (Harris 1996).

\subsection{ULX Candidate 2}

Only 18 photons were detected from the \cxo\ image of source \#~2
 corresponding to an intrinsic luminosity of 
 (1.0$\pm$0.4)~x~10$^{39}$~erg~s$^{-1}$.
Its X-ray colors are unremarkable (see Figure~3). 
A potential optical counterpart is present in the
\hst\ V and I images within 0.\arcsec10 of the \cxo\ position. 

\clearpage
\begin{center}
\includegraphics[angle=-90,width=\columnwidth]{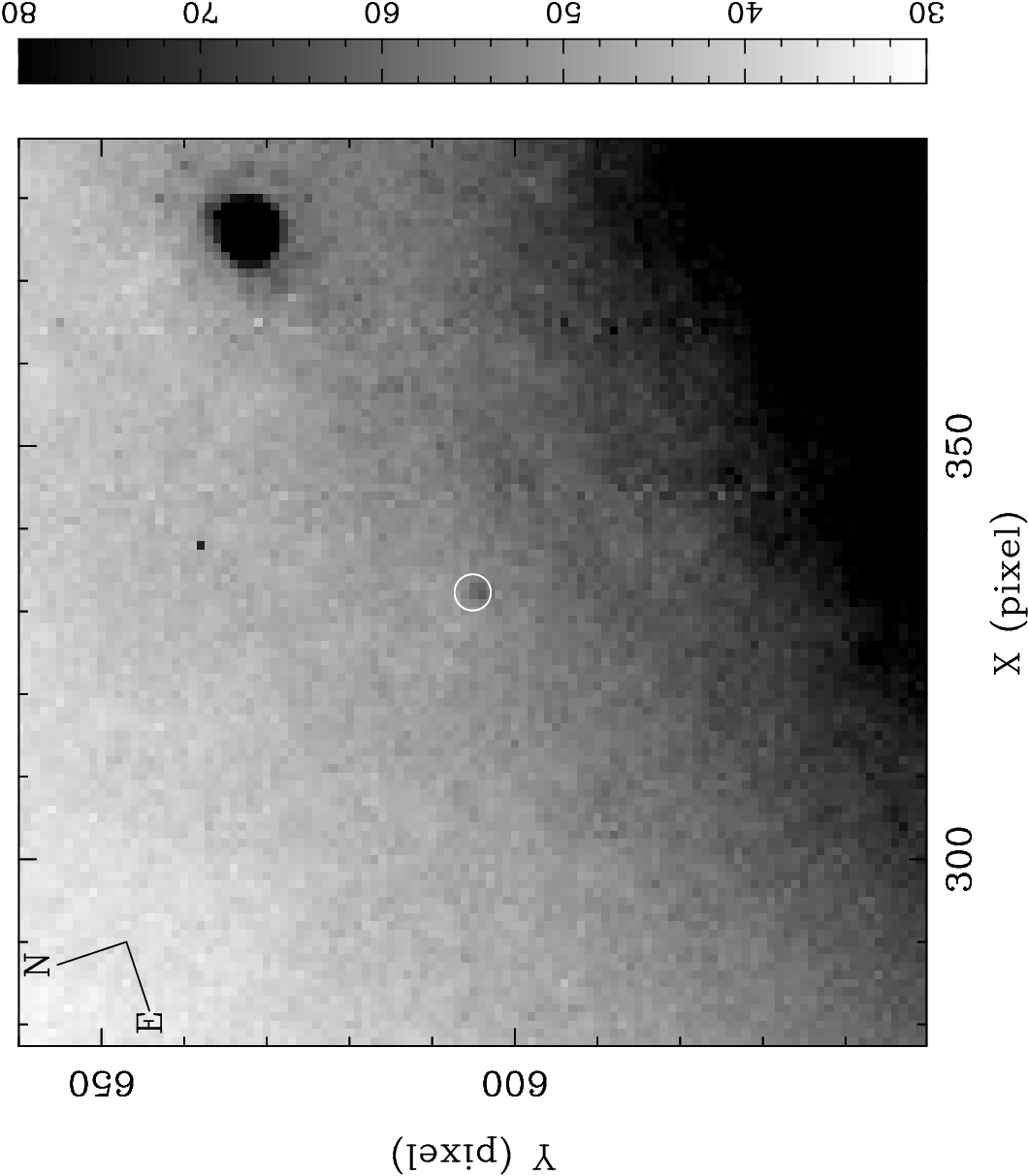}
\figcaption{
\hst\ 5\arcsec$\times$5\arcsec\ F814W band PC1 camera image 
(5240~s intergration) around ULX candidate \#~2.
After astrometric corrections, the positional offset
between the \hst--counterpart and the ULX candidate\#~2 is 0.\arcsec10. 
The circle depicts the 0.\arcsec1 radius \cxo\ error circle.
}
\end{center}
\clearpage

Figure~7 shows a portion of 
 the I band \hst\ image with the 0.\arcsec15 \cxo\ error circle superposed. 
The source is close to the bright nucleus and this accounts for the large 
 gradient visible in the field surrounding the source. 
This source was observed on two occasions with the HST/PC1 camera and no variability
was detected.
The M$_{V}$ and (V--I) values obtained for the counterpart 
 (Figure~4 and Table~3) are consistent with those 
 of Galactic globular clusters. 
Thus, source \#~2 could be a ULX within a globular cluster in
 NGC~5018.  

\subsection{ULX Candidate 3}
Source \#~3 is bright enough in X-ray light to constrain its spectral shape. 
The X-ray spectrum is that of a modestly-absorbed power law (Table~1)
 with no other distinguishing spectral characteristics 
 or unusual X-ray colors (Figure~3). 
\clearpage
\begin{center}
\includegraphics[angle=-90,width=\columnwidth]{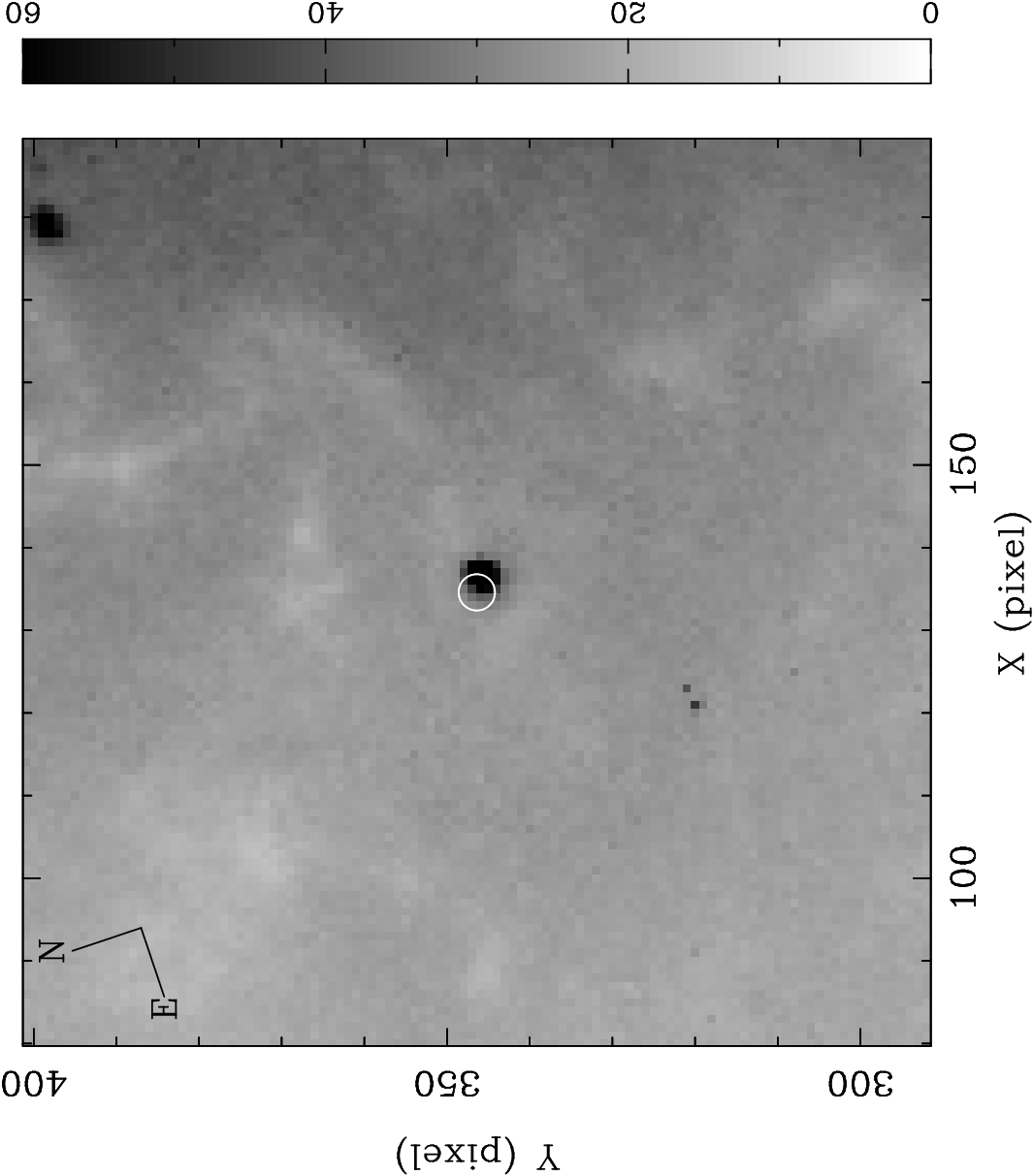}
\figcaption{5\arcsec$\times$5\arcsec\ region of the 4440 s \hst\ F555W band PC1
 image around the ULX candidate \#~3.
The \hst\ object is 0.\arcsec1 from center of
the \cxo\ position at the edge of the \cxo\ error circle.}
\end{center}
\clearpage

The optical object at the location of source \#~3 is the only counterpart 
 detected in all three \hst\ filters.  
Figure~8 shows the V band \hst\ image and Figure~9 the U band image. 
The region of NGC~5018 containing source \#~3 was observed twice in V and I
 bands and once in U (Table~2). 
Table~3 shows that the counterpart to source~\#~3 varied in both V and I by
 about 1~mag between observations.
Since the U band observations were not contemporaneous with either of the
 the V and I observations, we cannot make meaningful estimates of the (U--V)
 or (U--I) colors of the counterpart to source \#~3.

The (V--I) color indicates the object is on the blue side of typical globular
 clusters and bordering on the SSC region (Figure~4).
However, the variability is inconsistent with GC (or SSC) behavior.
The \fxfo\ ratio is of order unity (Table~3) whereas foreground
 stars have lower values (Maccacaro \etal\ 1988).
We conclude, therefore, that source \#~3 is most likely a background AGN.

\clearpage
\begin{center}
\includegraphics[angle=-90,width=\columnwidth]{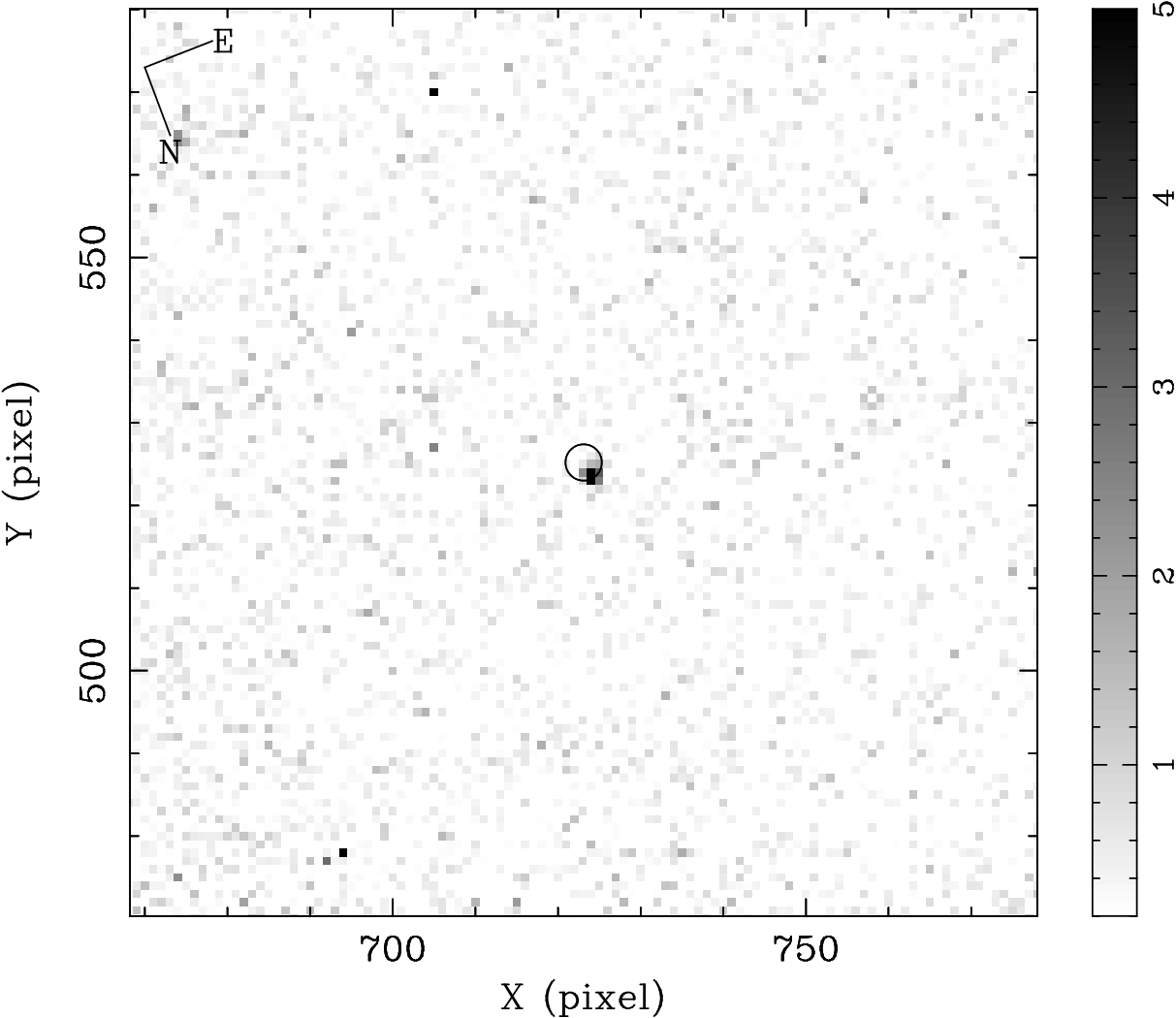}
\figcaption{5\arcsec$\times$5\arcsec\ region of the 1800 s \hst\ F336W band PC1
 image around the ULX candidate \#~3.
The \hst\ object is 0.\arcsec1 from center of
the \cxo\ position at the edge of the \cxo\ error circle.
}
\end{center}
\clearpage

\subsection{ULX Candidate 4}

Source \#~4 is another weak X-ray emitter with non-distinct X-ray colors.
No obvious optical counterpart can be discerned from either the available
 \hst\ V (Figure~10) or I band images of the field and only upper limits 
 to a potential counterpart are listed in Table~3. 
In this regard, it is similar to source~\#~1 though source \#~1 is much
 brighter in X-rays than source \#~4.

\clearpage
\begin{center}
\includegraphics[angle=-90,width=\columnwidth]{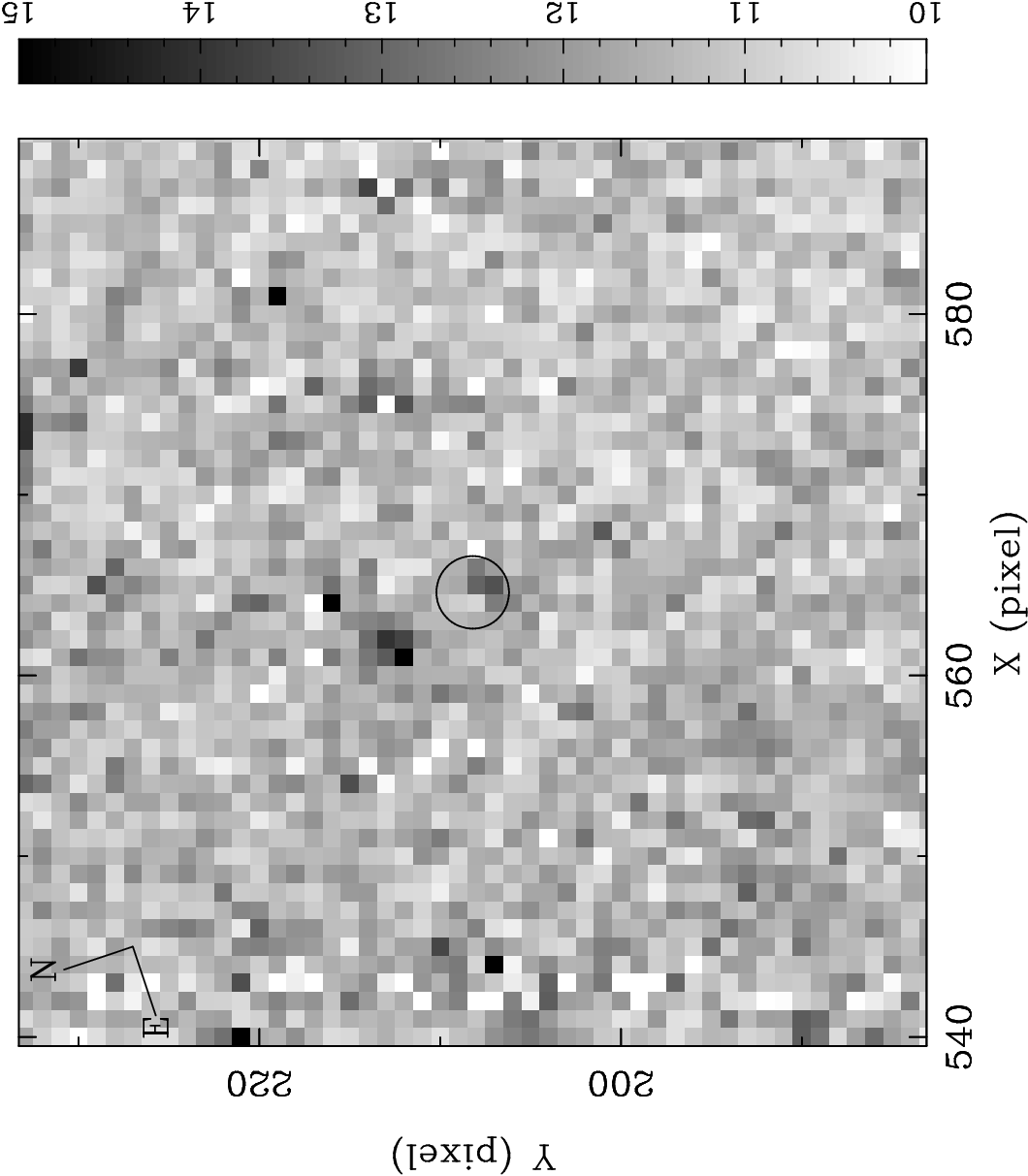}
\figcaption{5\arcsec$\times$5\arcsec, 4440~s \hst\ F555W image around 
ULX candidate \#~4 with the 0.20\arcsec\ radius \cxo\ error circle added. 
There is about a 2.5 percent chance of finding a source as bright as
the one in the error circle.}
\end{center}
\clearpage

\subsection{ULX Candidate 5}

Source \#~5 is moderately weak in X-rays with 42 counts detected. 
This corresponds to an intrinsic X-ray luminosity of 
 (2.4$\pm$0.3)$\times 10^{39}$~\ergl. 
A potential counterpart was detected in both V and I images. 
After applying astrometric corrections, the positional difference between the 
counterpart and the ULX candidate is 0.\arcsec20.
Figure~11 shows the V band \hst\ image in the vicinity of ULX candidate \#~5.  
The location of the counterpart of ULX \#~5 in Figure 4 suggests that it may be either a GC or a SSC in NGC~5018 or a background AGN. 

\clearpage
\begin{center}
\includegraphics[angle=-90,width=\columnwidth]{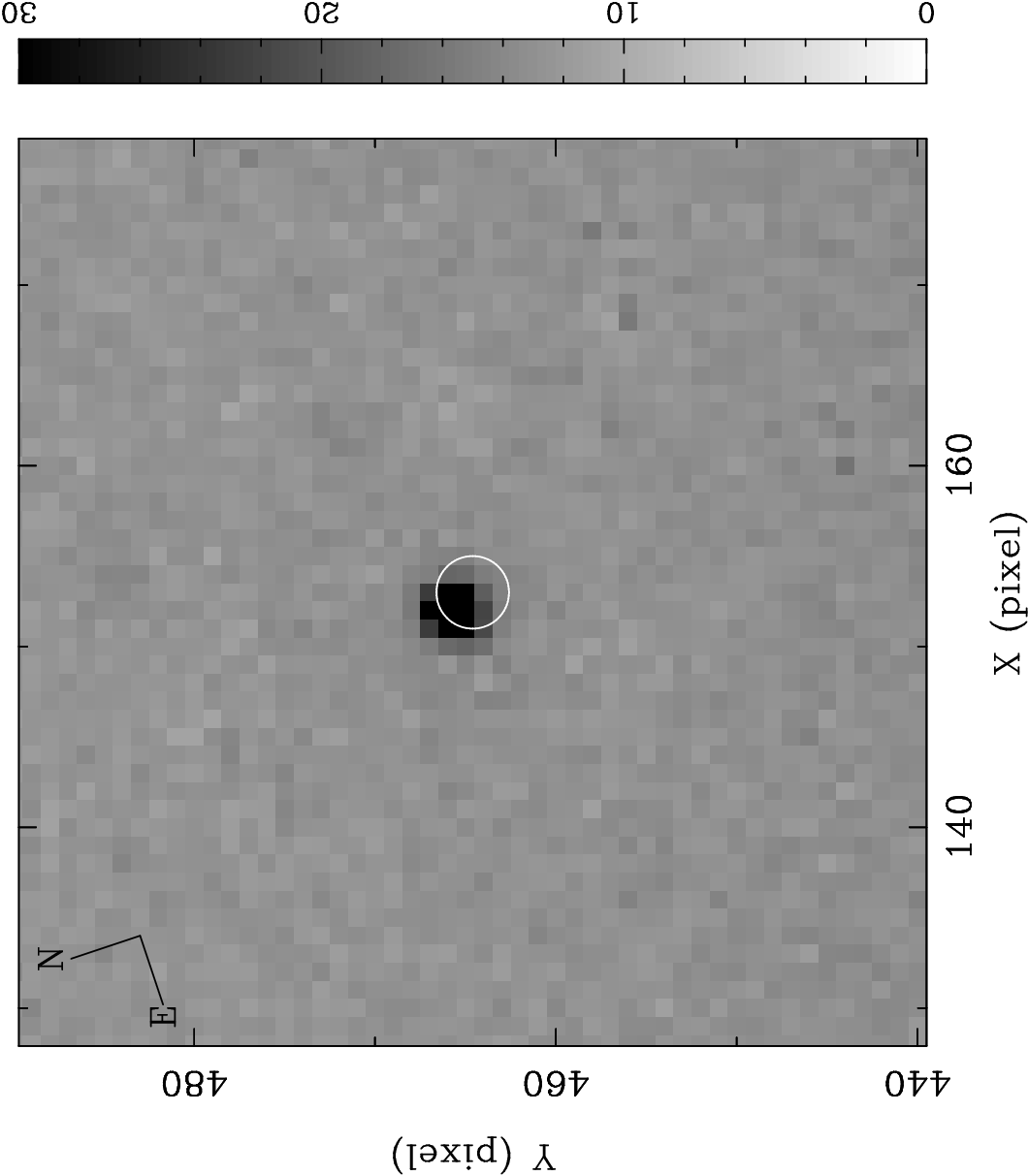}
\figcaption{5\arcsec$\times$5\arcsec\ F555W band image around the position
of ULX candidate \#~5.
The 0.\arcsec 2 \cxo\ error circle encompasses a bright \hst\ source. 
}
\end{center}
\clearpage

\subsection{ULX Candidate 6}

Source \#~6 is a weak X-ray source but is distinguished by an extremely
 hard X-ray spectrum as evidenced by its position in the X-ray color-color
 diagram (Figure~3).
Although the X-ray colors suggest a very flat spectrum, the low S/N data does not allow us to rule out a more-typical power law index of $\Gamma$$\sim$1.8. 
The region around this source was imaged only in the V and I filters.
After astrometric corrections were applied to these images, a bright source
 was detected within 0.\arcsec2 from the \cxo\ location (Figure~12).
The optical source is similar to the other bright ULX counterparts in that
 it has a color and magnitude consistent with a globular cluster in NGC~5018.

\clearpage
\begin{center}
\includegraphics[angle=-90,width=\columnwidth]{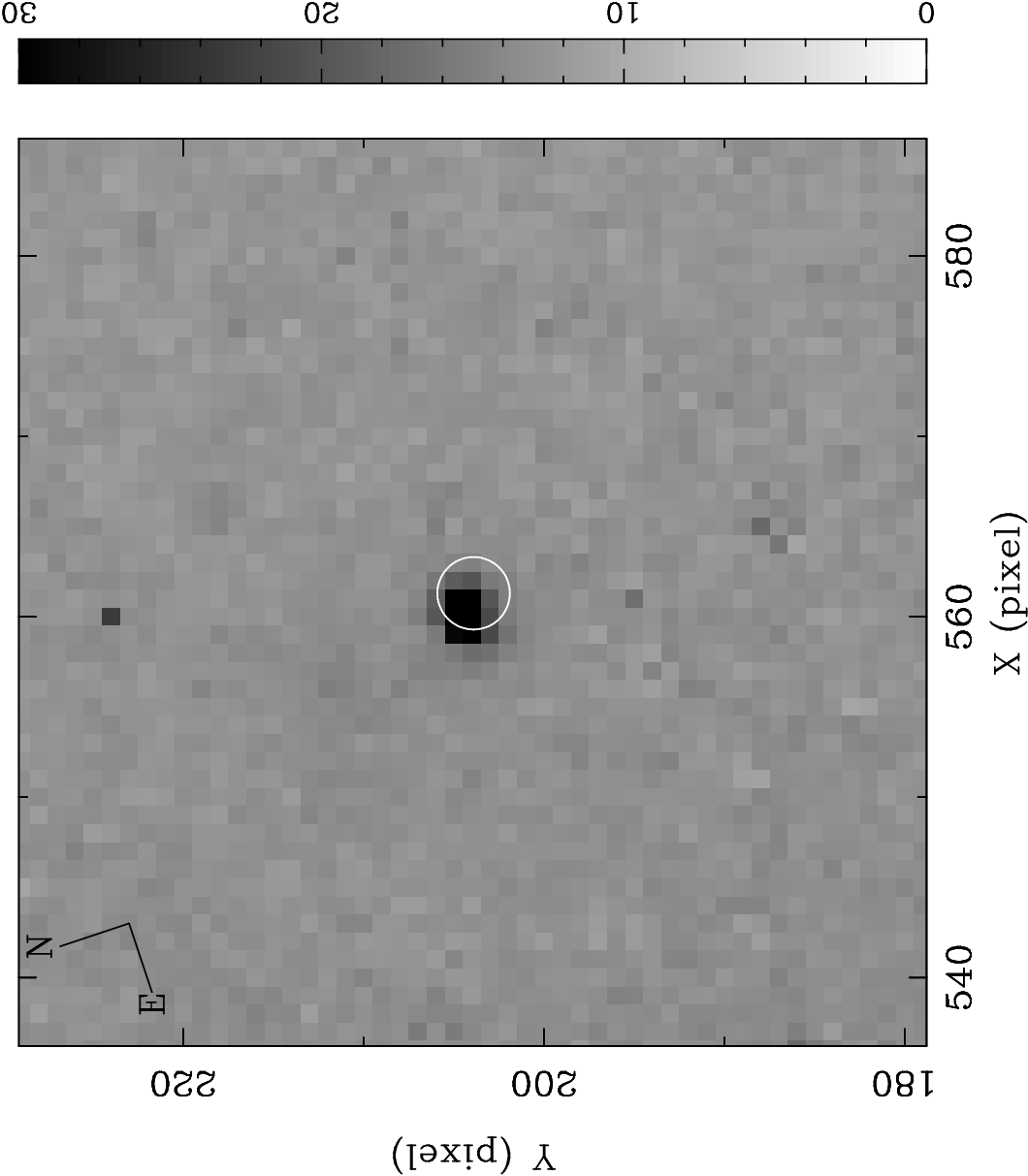}
\figcaption{5\arcsec$\times$5\arcsec\ F555W band image around the position
of ULX candidate \#~6.
The 0.\arcsec 2 \cxo\ error circle encompasses a bright \hst\ source. }
\end{center}
\clearpage

Unlike all other ULX candidates in the NGC~5018 field, 
 high-resolution radio data confirms (Mollenhoff \etal\ 1992) the 
 detection of a radio source (Kim \etal\ 1988) at the position of 
 source \#~6 (see Table~5). 
After applying astrometric corrections based on the X-ray and radio
 positions of the nucleus, the position of the radio source
 is within 0.\arcsec15 of the \cxo\ position of source~\#~6. 
This was confirmed through our multi-wavelength radio observations obtained
 with the Australia Telescope Compact Array (ATCA). 
We obtained radio continuum data in total and polarized intensity on June 30, 
 2004, simultaneously observing at four frequencies,
 1.4, 2.3, 4.8, and 8.6 GHz with the ATCA in the 1.5B array. 
Measured flux densities are given in Table~5.
The radio spectrum is inverted with a spectral index of $\sim$~0.2.
The source shows no polarization above 3 sigma at any of the four frequencies. 

One model for ULXs invokes relativistic beaming to account for the high
 apparent X-ray luminosity (Georganopoulos \etal\ 2002).
The radio flux should also be high in this scenario.
However, the X-ray spectra 
 of Galactic and extragalactic jet sources are typically 
 much steeper than that of source \#~6 and the observed 
 inverted radio spectrum is not typical of jets which have 
 radio spectral indices near -0.6 (Worrall \& Birkinshaw 2004).
On the other hand,
 an inverted spectrum and a lack of polarization are indicative of 
 compact cores of radio galaxies.


\section{The Nuclear Region}

The nucleus of NGC~5018 is surrounded by an $\sim$15\arcsec\ radius
 ($\sim$3 kpc) region of unresolved emission.
Figure 13 displays the radial profile of the X-ray surface brightness about the
 nucleus.
Also shown is the radial profile of a nearby point source (ULX candidate \#3)
 clearly demonstrating the extended nature of the nuclear emission.

\clearpage
\begin{center}
\includegraphics[angle=-90,width=\columnwidth]{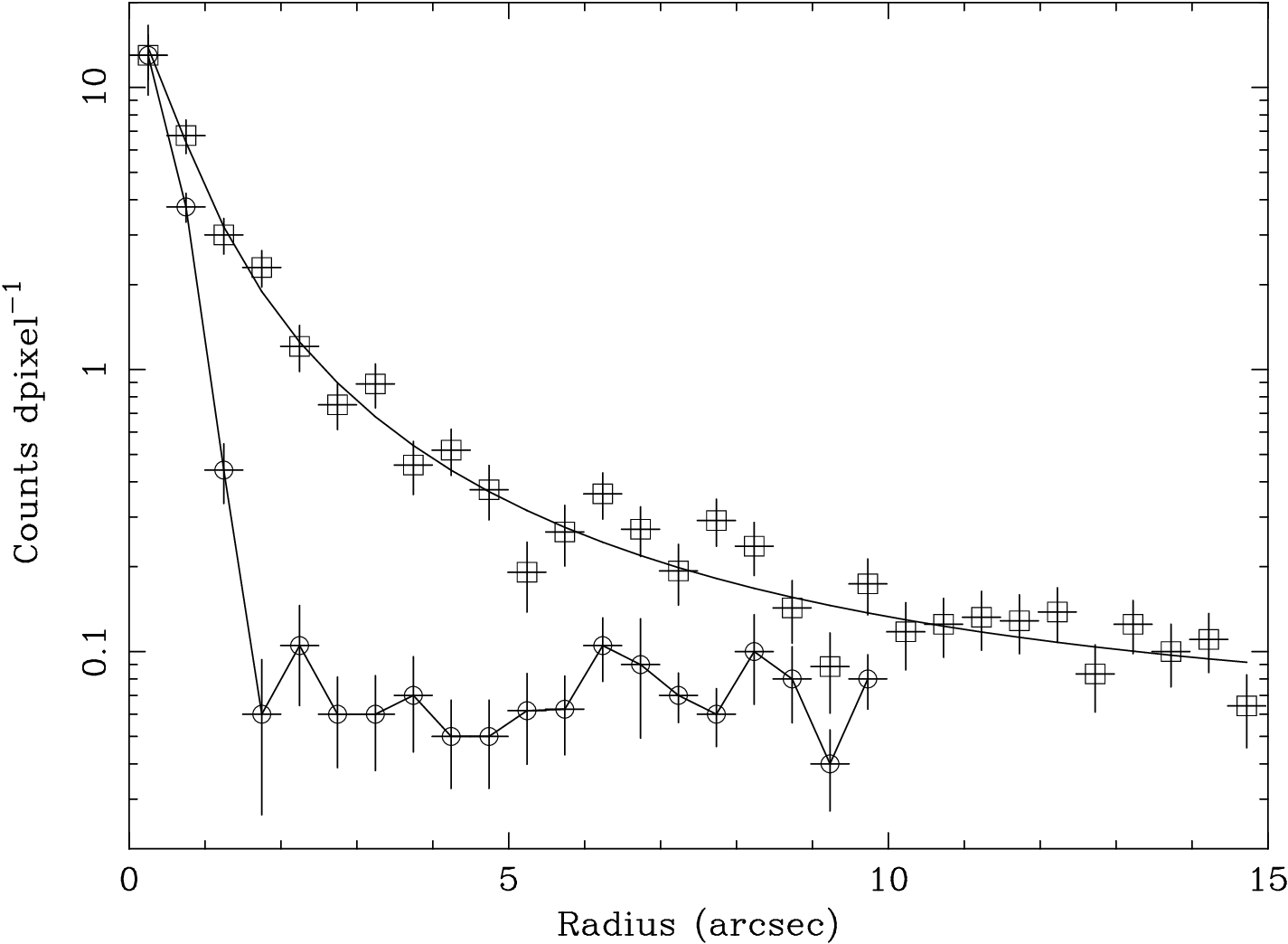}
\figcaption{
Radial profile of the full band (0.5--8.0~keV) X-ray surface brightness
 about the nucleus of NGC~5018 (open squares).
The radial profile of ULX candidate \#~3, located 15\arcsec\ from the nucleus,
 is also shown for comparison (open circles), which has been scaled to match at the peak. 
The smooth solid curve represents a standard beta model profile
 as discussed in the text.
}
\end{center}
\clearpage

If the nuclear emission is from isothermal gas in hydrostatic equilibrium,
 then the surface brightness profile would follow a standard beta model,
 $\propto$$(1+(r/r_c)^2)^{-3\beta+1/2}$.
Fitting this function (plus a constant representing the background)
 to the 0.3--8.0~keV X-ray profile
  out to 15\arcsec\ from the nucleus results in values of
 $r_c = 0.\arcsec 46\pm0.\arcsec 17$ ($\sim$90 pc), and
 $\beta = 0.44 \pm 0.02$ ($\chi^2$=34.0 for 27 dof).
This function is shown as the smooth curve in Figure~13.
The small core radius, $r_c$, and the low value of $\beta$
 means the profile forms a cusp at the center but is rather flat at larger
 radii.

The cusp may be due to a true point source at the nucleus;
 consistent with the identification of a point-like object at that location
 by our source-finding algorithm.
This source has a soft spectrum as shown by its X-ray colors
 (\S~3.1), but a hard spectral component is often also present in
 active galactic nuclei.
Therefore, the radial profile in the 2--8~keV band was extracted and compared
 to the full-band profile.
The hard-band profile appears more concentrated
 toward the nucleus (best-fit $r_c=0.\arcsec 25 \pm 0.\arcsec 18$,
 $\beta = 0.46 \pm 0.04$; $\chi^2$ = 32.3 for
 27 dof) compared with the full band profile but is consistent with that
 profile within errors.
It cannot be concluded, based on the surface brightness profiles alone,
 whether or not there is an X-ray-emitting active nucleus in NGC~5018.

Guided by the known angular size of a point source at the position of the
 nucleus, we extracted the X-ray spectrum from two regions:
 One from an annulus extending from 3\arcsec\ to 20\arcsec\ around the
 nucleus and excluding the point sources within this region
 (henceforth, the extended component), and the other
 from a disk of radius 3\arcsec (the nuclear component).
The 20\arcsec\ outer radius was chosen
 to roughly correspond to the (spatially-uniform) background level.
The spectra were binned to obtain at least 20 counts per fitting bin.
A corresponding background spectrum was extracted from a circular annulus that
 extends from 40\arcsec\ to 70\arcsec\ radii, again excluding point sources.

Various XSPEC models were fitted to the extended component spectrum as
 presented in Table 4 (labeled ``Diffuse'' in column 1).
The spectrum requires at least two spectral components for an acceptable fit.
The best fit is a thermal emission line ({\tt mekal}) plus
 power law model (row 6, Table~4).
The abundances of metals is poorly constrained by this model but are
 consistent with solar values.
The spectrum and best-fit model are shown in Figure~14.
Extrapolating the luminosity in the 3\arcsec\ to 20\arcsec\ annular
 spectral extraction region to the center of the galaxy using the beta model
 shape
 results in a total 0.5--8.0~keV absorption-corrected luminosity of $
 (1.62\pm0.86) \times 10^{40}$~\ergl.

\clearpage
\begin{center}
\includegraphics[angle=-90,width=\columnwidth]{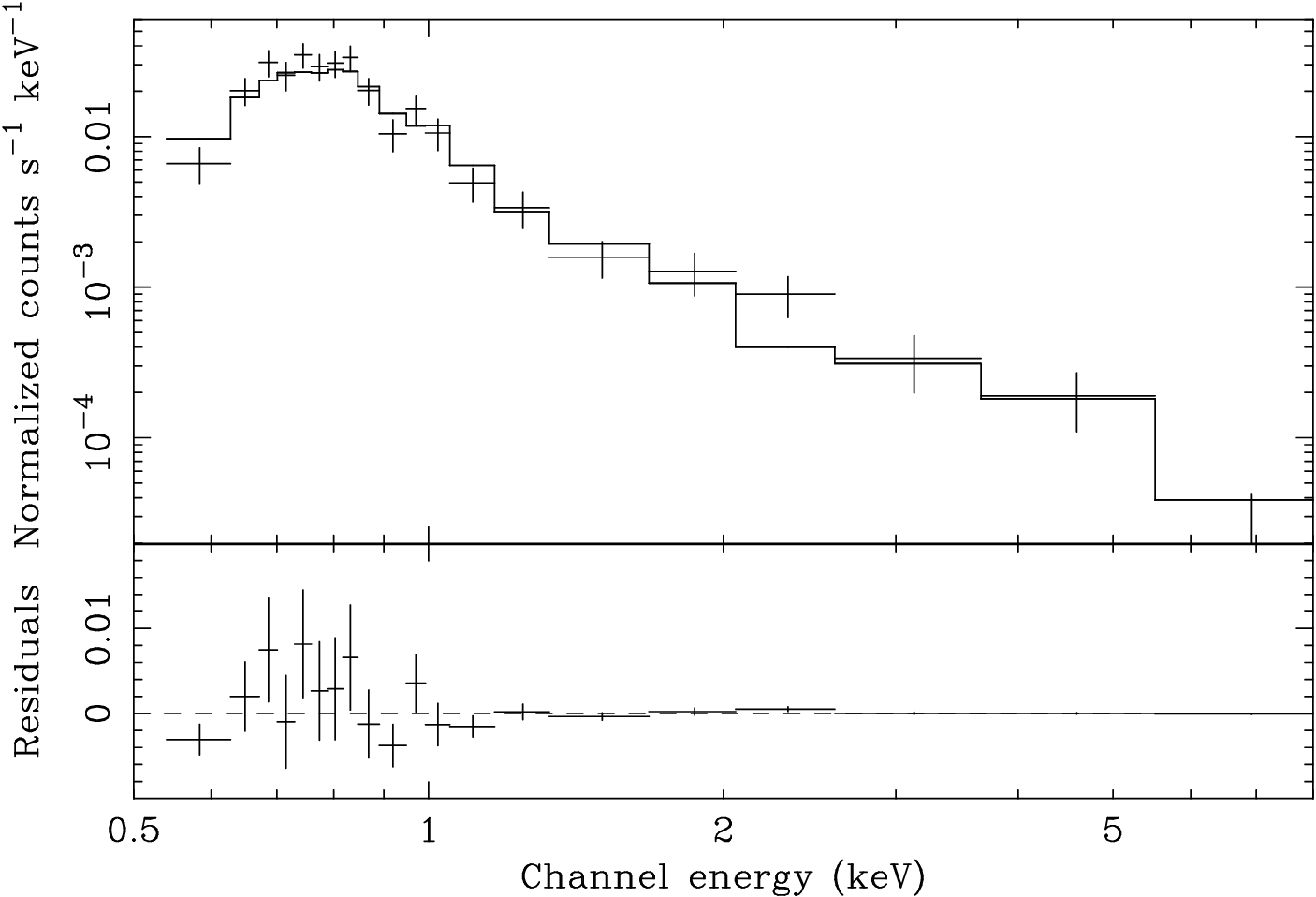}
\figcaption{
X-ray spectrum of the extended, diffuse
 emission around the nucleus of NGC~5018,
 extracted from an annulus with 3\arcsec\ and 20\arcsec\ radii (upper panel).
Also shown are the best-fit model composed of
 absorbed Mekal and power law components.
Fit residuals are shown in the lower panel.
}
\end{center}
\clearpage

The shape of the extended X-ray spectrum
 suggests a hot gas origin for the X-ray emission.
This is supported by observations at other wavelengths.
H$\alpha$ emission has been detected extending to
 32\arcsec\ ($\sim$ 6.3 kpc) from the nucleus with a total luminosity
 $\sim (6.8\pm0.4) \times 10^{39}$~\ergl\ (Goudfrooij et al. 1994).
This region also has a distinctly bluer color
($\Delta(B-V) \sim 0.12$~mag) relative to its surroundings (Carollo \& Danziger 1994).
A portion of the radio emission can also be attributed to an extended source.
The total 20~cm continuum emission measured over a 20\arcsec\ region around the nucleus is
3.1$\pm$0.17 mJy and that from the nucleus itself ($<$1\arcsec) is (1.9$\pm$0.1) mJy 
(Mollenhoff \etal\ 1992).
Thus, about 1.2$\pm$0.2 mJy is attributable to the extended emission.
The radio luminosity is then $\sim (3.6\pm0.6) \times 10^{36}$~\ergl.
This is consistent with our recent radio measurements (Table~5).

Some of the extended X-ray emission may also be due to unresolved point sources.
In particular, the hard
 (power law) spectral component may be produced by
 LMXBs in the relatively dense core of NGC~5018.
The fraction of the extended X-ray emission attributed to this component is
 0.46 for a luminosity of $\sim (5.9\pm2.9) \times 10^{39}$~\ergl.
If we assume that of order 10$^{37}$~\ergl\ is an average X-ray luminosity
of LMXBs,
 then $\sim$600 unresolved LMXBs are required to account for the
 hard X-ray emission from the extended component.
This is not an unrealistic number of LMXBs in a typical elliptical galaxy.
However, a population of LMXBs cannot account for the extended emission
 observed at other wavelengths.
Measured radio luminosities of Galactic LMXBs (e.g., Han  \& Hjellming 1992;
 Berendsen et al. 2000; Homan et al. 2004) are only of order
 $\sim$10$^{32}$~\ergl\ and Balmer line emission from LMXB accretion disks
 is of order a few 10$^{32}$~\ergl.
Based on these results, we believe that the extended radio, H$\alpha$,
 and at least 54\% of the X-ray emission are from diffuse hot gas in the
 central regions of NGC~5018.

The spectrum of the nuclear component is also best represented by a
 two component model (row 8 in Table~4; Figure~15).
Of course, we expect a soft component analogous to the soft emission
 from the surrounding region.
Unfortunately, the hard component is also consistent with an extrapolation
 of the power law emission from beyond the nucleus and so, again, we cannot
 unequivocally confirm an active nucleus for NGC~5018.

If we assume a nuclear point source is present, then we can compare it 
 to other low luminosity AGNs (LLAGNs). 
Within 3\arcsec\, the power law component contributes 51\% of the total 
 luminosity or L$_{X}$~=~(3.5$\pm0.7) \times 10^{39}$~\ergl.
We consider this as the maximum luminosity from any central point source. 
To obtain a minimum luminosity estimate, we
performed a joint fit to the nuclear and extended spectra 
using a mekal plus powerlaw model. Then, with the constraint that the same parameters are used to fit
both spectra, a second power law component
was added to the model of the nuclear spectrum only. 
The 90\% confidence upper limit to the contribution from the second powerlaw component is L$_{X}$=2$\times 10^{37}$~\ergl\ with a best-fit value
of zero.
Using these X-ray luminosity limits,
the resulting ratio of the 5 GHz flux density (Table~5) to the X-ray 
 luminosity, log[$\nu L_{\nu}$(5GHz)/L$_{X}$], is between -2.4 and -0.15.
This is within the range of -4.6 to -1.2 typical of LLAGNs
 (Terashima \& Wilson 2003).
In addition, the value of log(L$_{X}$/L$_{H\alpha}$) for NGC~5018 is between 
 -1.5 and 0.74 assuming all the H$\alpha$ emission from the 3\arcsec\ 
 nuclear region is associated with the point source at the nucleus. 
Terashima \& Wilson (2003) find log(L$_{X}$/L$_{H\alpha}$) for LLAGNs are in 
 the range -0.6 to 2.0.

\clearpage
\begin{center}
\includegraphics[angle=-90,width=\columnwidth]{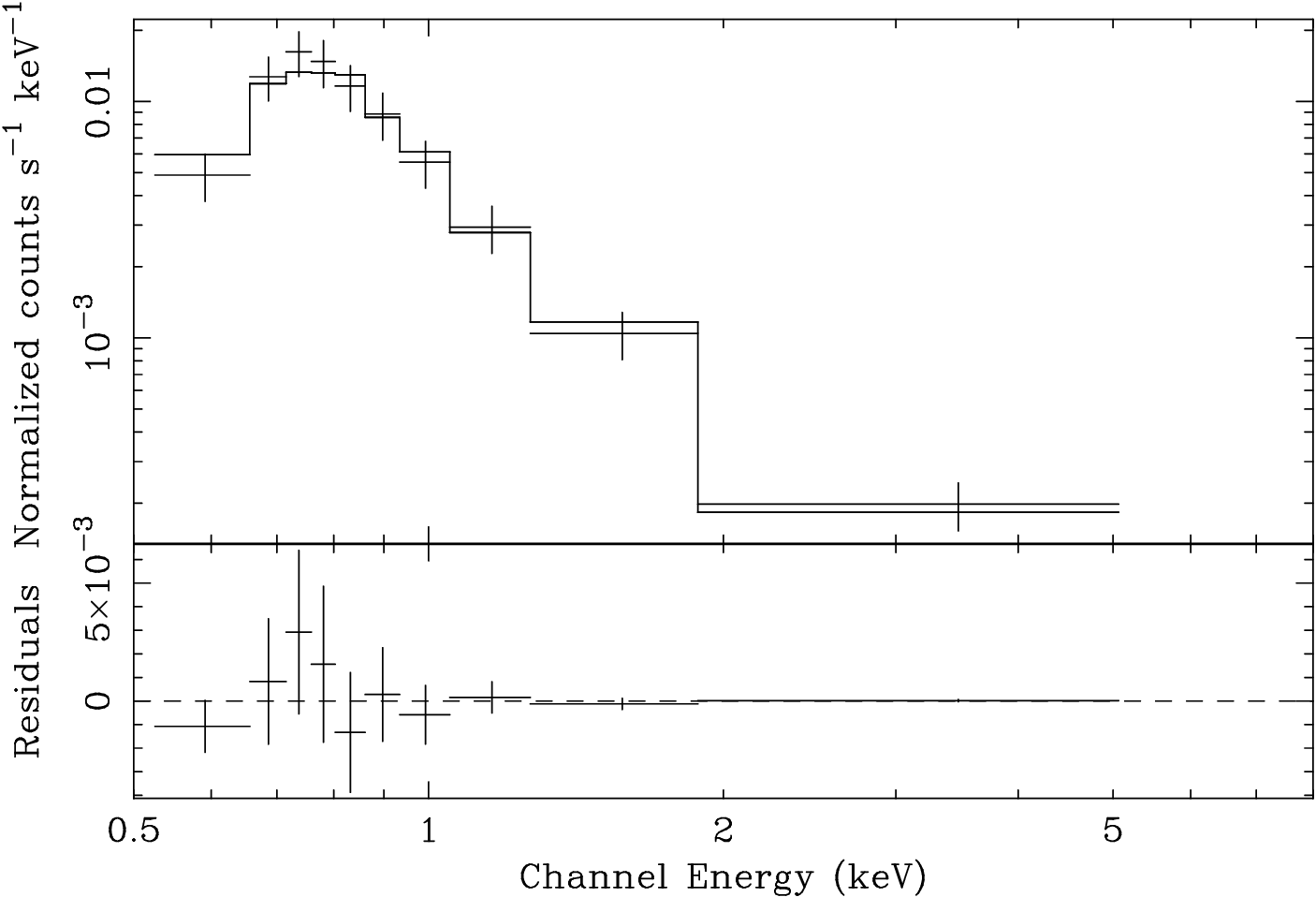}
\figcaption{X-ray spectrum of the emission within a
3\arcsec\ radius of the center of NGC~5018. This spectrum is best fitted with an absorbed 
thermal plus power-law model. Data and the model are shown in the upper panel and the 
residual between the data and the model is shown in the lower panel of this figure.}
\end{center}
\clearpage

There is no evidence for variability of the nuclear emission during the
30~ks \cxo\ observation. The light curve of the hard X-ray emission, $>$2~keV,
is of insufficient quality to determine if the power-law emission
component alone is variable.

\section{Summary and Discussion}

Six non-nuclear X-ray point sources were detected in NGC~5018 in a 
 30~ks \cxo\ observation.
All six sources are potentially ULXs and each displays interesting X-ray and
 optical properties.
An estimated 1.4 sources are background objects. 

The dominant contribution to the X-ray point source population in early-type
 galaxies is LMXBs. 
Individually, accretion disks and companion stars of 
 LMXBs are too faint at non-X-ray wavelengths to be detectable
 at the distance of NGC~5018 even with \hst.
Sources \#~1 and \#~4 are in this category.
Only upper limits to potential optical counterparts to these sources 
 could be determined and this limit, $M_V$\LA -6.5, excludes only the 
 brightest early-type supergiant companions.
A better constraint on the companion mass can be made if
 mass transfer is assumed to be
 driven by the nuclear evolution of the companion. 
In this case, the observed X-ray luminosities can be used to constrain 
 the mass transfer rate and hence the companion mass.
If the companion is a main sequence star, then it must be more massive
 than 8 and 20 \msun\ for source \#~4 and \#~1, respectively
 (assuming a 10\% efficiency converting accreting matter to X-radiation).
If the companion is in its giant phase, then these values are reduced to
 4 and 8 \msun. 
(Note that these two sources span the range of X-ray luminosities of all the
 non-nuclear sources so the stellar companions to these other sources also have
 masses in the range 4 to 20 \msun\ by these arguments.)
Stars of such high mass are too short-lived to have been formed at the
 time of interaction between NGC~5018 and other group member galaxies.
Perhaps, as discussed by King (2002), these objects are soft X-ray 
 transients in outburst with emission beamed in the direction of the
 observer. 
Alternatively, Bildsten \& Deloye (2004) propose ultracompact binaries
 that can produce very high mass accretion rates from gravitational
 radiation losses and hence X-ray luminosities in excess of 
 $\sim$5$\times$$10^{38}$~\ergl.
 
In any case, the remarkably high X-ray luminosity of source \#~1,
 $>$$10^{40}$~\ergl\ based on spectral 
 fitting\footnote{The X-ray luminosity of source \#~1 exceeds that of all 
 57 ULX candidates detected in 27 elliptical galaxies surveyed by 
 Swartz \etal\ (2004)}, is perplexing. 
No model of LMXBs can easily account for such a high X-ray luminosity.
Statistically, background objects are most likely to be among the 
 brightest X-ray objects in elliptical galaxy fields (Irwin, Bregman, \& Athey  2004). 
On the other hand, background AGNs tend to have X-ray to optical flux
 ratios of order unity (e.g., Maccacaro \etal\ 1988; Green \etal\ 2004)
 predicting an optical counterpart to source \#~1
 far above the \hst\ detection limit, though none was found for source \#~1 
 (ratio $\ge$186; Table~3). 

The remaining four sources are all associated with optically bright objects.
The (V--I) colors and V magnitudes of these objects are consistent with 
 GCs in NGC~5018 (Hilker \& Kissler-Patig 1996).
However,  
 source \#~3 is likely a background AGN based on its optical variability and
 source \#~6 is likely a background radio galaxy based on its radio
 signature.
Without a spectroscopic redshift confirmation, we cannot be certain if
 any of these four sources are truly ULXs in NGC~5018 GCs or background objects.
Statistically, we expect 1.4 background objects among our 6 ULX candidates,
 and we expect the X-ray luminosities of the true ULXs to be below 
 $\sim$2$\times$$10^{39}$~\ergl\ (Irwin, Athey, \& Bregman 2003; Irwin, Bregman, \& Athey  2004).
Statistically, therefore, sources \#~3 and~6 can account for the expected 
 background contribution.
In the broader context, we caution that X-ray bright background objects also
tend to be optically bright.

In addition to point sources and a possible weak LLAGN, diffuse hot gas has also
 been detected in the Xradiation from NGC~5018.
This gas may be the remnant of interactions of NGC~5018 with its neighbor
 galaxies. 
The radiative cooling time of the hot plasma, estimated from the 
 X-ray spectral fit parameters and the size of the emission region, is
 a few times 10$^{7}$ to 10$^{8}$~yr, depending on the gas filling factor.
This is shorter than the estimated 300 to 600~Myr since the last 
interaction (Guhathakurta \etal\ 1989). 
This suggests the gas is still falling in and/or it is being reheated by 
 ionization, stellar winds, and supernovae from current or 
 recent star formation activity.
The current star formation rate (SFR) 
 implied by the H$\alpha$ luminosity is only
 $\sim$0.1~\msun\ yr$^{-1}$ but some of this gas is obscured by overlying
 dust (Goudfrooij \etal\ 1994).
The far-infrared luminosity of NGC~5018 is 
 2$\times$10$^{43}$~\ergl\ (Thronson \& Bally 1987)
 corresponding to a star formation rate of $\sim$0.9 \msun\ yr$^{-1}$.
This, in turn, may be an overestimate as there is a substantial
 contribution to the 
 FIR luminosity from dust heated by the interstellar radiation field
 (Guhathakurta \etal\ 1989) in addition to the warm component 
 normally associated with young star-forming regions.
The mass of atomic and molecular gas in NGC~5018 is 
 $M_{gas}$$\sim$1.4$\times$$10^9$~\msun\
 (Georgakakis, Forbes, \& Norris 2000) implying a gas consumption timescale
 ($M_{gas}$/SFR) of order a Hubble time.
Thus, while there is little current star formation activity,
 there is a significant reservoir of gas to maintain  
 this low, steady, level of star formation since the last interaction that
 may explain the diffuse thermal X-ray (and H$\alpha$ and radio) emission 
 from the central regions of NGC~5018.

We may obtain a broad, though speculative, interpretation of the recent 
 history of NGC~5018 from its X-ray to radio properties.
The morphology and radial velocity structure of the HI gas indicate
 the interaction between NGC~5018 and NGC~5022 was not a direct impact and
 merger but an encounter that left the more massive NGC~5018 essentially intact.
The interaction did strip gas from NGC~5022 that presently traces
 the path of the interaction in HI (Kim \etal\ 1988; Guhathakurta \etal\ 1989),
 produced the optical shells (Malin \& Hadley 1997), and is 
 currently accreting toward the center of NGC~5018 where it is being heated
 and ionized.
Stars are currently forming out of this gas at a low rate but the rate may
 have been higher in the recent past if the ULX sources are binaries accreting 
 from normal or giant stars. 
Within the stellar system is the
 usual population of LMXBs. Most lie below the point source detection limit 
 in the current \cxo\ data but a number of exceptionally luminous (ULX)
 candidates are present. These appear to trace the relatively young population
 of stars contained in blue GCs that may have formed since the last interaction.

\acknowledgements
Our sincere thanks to the Editor and the referee for their comments that helped to 
improve the quality of the figures.
This research has made use of the NASA/IPAC Extragalactic Database (NED) which
 is operated by the Jet Propulsion Laboratory, California Institute of
 Technology, under contract with NASA;
of data products from the Two Micron All Sky Survey, which is a joint project 
 of the University of Massachusetts and the Infrared Processing and Analysis 
 Center, funded by NASA and the NSF;
from the Multimission Archive (MAST) at the STScI operated by AURA under NASA 
 contract NAS5-26555;
and from the Chandra Data Archive, part of the Chandra X-Ray Observatory  
 Science Center (CXC) which is operated for NASA by SAO. The Australia Telescope is 
funded by the Commonwealth of Australia for operation as a National Facility managed
by CSIRO.
Support for this research was provided in part by NASA under 
 Grant NNG04GC86G issued through the Office of Space Science.

\vspace{15pt}

\begin{center}
\scriptsize{
\begin{tabular}{cccccccccc}
\multicolumn{10}{c}{{\sc Table 1}} \\
\multicolumn{10}{c}{NGC 5018 X-Ray Source Properties} \\

\hline \hline

 &  \multicolumn{1}{c}{R.A.} & \multicolumn{1}{c}{Dec.}  & \multicolumn{3}{c}{Spectral Parameters$^a$} & \multicolumn{1}{c}{$L_X^b$} & \multicolumn{1}{c}{$(M-S)/T^c$} & \multicolumn{1}{c}{$(H-M)/T^c$} & \multicolumn{1}{c}{Detected}  \\
 &  \multicolumn{1}{c}{(J2000)} & \multicolumn{1}{c}{(J2000)}  & $N_H$ & $\Gamma/kT_e$ & $\chi^2$/dof & \multicolumn{1}{c}{($10^{39}$~erg~s$^{-1}$)}
  & & & \multicolumn{1}{c}{Counts,$T$}\\
\hline
1 & 13 12 55.603 &-19 30 39.60  & $8.7^{+12.2}_{-8.7}$& $1.74^{+0.43}_{-0.37}$& 2.1/8 & 12.9$\pm$1.7 & +0.15$\pm$0.06   & -0.19$\pm$0.06 &233 \\
2 & 13 13 01.200 &-19 30 57.14  & ---& --- & --- & 01.0$\pm$0.4 & +0.27$\pm$0.17 & -0.13$\pm$0.04 & 18 \\
3 & 13 13 02.036 &-19 31 05.44  & $20.4^{+25.5}_{-20.4}$& $1.26^{+1.1}_{-0.4}$& 3.5/2 & 06.3$\pm$0.9 & +0.26$\pm$0.07 & +0.07$\pm$0.08 &121 \\
4 & 13 13 02.708 &-19 32 11.56  & ---& --- & --- & 01.1$\pm$0.4 & +0.00$\pm$0.00 & -0.25$\pm$0.16 & 19 \\
5 & 13 13 04.343 &-19 31 52.62  & ---& --- & --- & 02.4$\pm$0.3 & +0.21$\pm$0.19 & -0.21$\pm$0.19 & 42 \\
6 & 13 13 06.476 &-19 31 14.82  & ---& --- & --- & 01.8$\pm$0.4 & -0.16$\pm$0.06 & +0.65$\pm$0.18 & 31 \\
7$^d$ & 13 13 01.048 &-19 31 05.60  & 6.98& $1.56^{+0.46}_{-0.59}$&3.8/7$^d$& 6.8$\pm$0.9 &-0.70$\pm$0.15 & +0.05$\pm$0.04 & 115\\
\hline

\multicolumn{10}{l}{$^a$Absorption column density unit is $10^{20}$ cm$^{-2}$; power-law photon index, $\Gamma$, given
for source 1; } \\
\multicolumn{10}{l}{\hspace{5pt}disk blackbody temperature, $kT_e$, for source 3.} \\
\multicolumn{10}{l}{$^b$Intrinsic 0.5--8.0 keV luminosity.}\\
\multicolumn{10}{l}{$^c$Colors from $S$ (0.5-1.0 keV), $M$ (1.0-2.0 keV) and $H$ (2.0-8.0 keV) bands.}\\
\multicolumn{10}{l}{$^d$Nucleus. Total counts from the 3$\arcsec$ region is 226. Mekal temperature is 0.33$\pm$0.1 keV.}\\
\end{tabular}
} 
\end{center}

\vspace{15pt}

\begin{center}
\small{
\begin{tabular}{cccccccc}
\multicolumn{8}{c}{{\sc Table 2}} \\
\multicolumn{8}{c}{\hst\ observations of NGC~5018} \\

\hline \hline

\multicolumn{1}{c}{Date of} & \multicolumn{1}{c}{Dataset} & \multicolumn{1}{c}{File name} & \multicolumn{1}{c}{Filter} & \multicolumn{1}{c}{Exposure (s)} & \multicolumn{2}{c}{Offset} & ULXs$^a$\\ 
\multicolumn{1}{c}{observations} & \#&&&&$\Delta\alpha(^s)$&$\Delta\delta(\arcsec)$&\\
\hline
7-29-1995 & 1& U2ST0201B & F336W  & 1800 & 0.018 & 0.6 & 1,2,3\\
7-28-1997 & 2 & U3M72505B & F555W &  1200 & 0.002 & 0.4 & 1,2,3\\
2-04-1999 & 3 & U5710401B  & F555W &  4440 & 0.028 & 0.5 & 2,3,4,5,6\\
7-28-1997 & 4& U3M72508B & F814W  & 1900 & 0.022 & 0.5 & 1,2,3\\
2-04-1999 & 5 & U5710407B &  F814W  &  5240 & 0.042 & 0.1 & 2,3,4,5,6\\

\hline
\multicolumn{8}{l}{$^a$Positions of ULXs covered in the image.}\\
\end{tabular}
} 
\end{center}

\begin{center}
\small{
\begin{tabular}{ccccccccc}
\multicolumn{9}{c}{{\sc Table 3}} \\
\multicolumn{9}{c}{The measured photometric magnitudes of counterparts of 6 ULX candidates in NGC~5018} \\

\hline \hline

\multicolumn{1}{c}{ULX} & \multicolumn{1}{c}{HST data-}&\multicolumn{1}{c}{Chandra error} & \multicolumn{1}{c}{F336W$^a$} & \multicolumn{1}{c}{F555W$^a$} & \multicolumn{1}{c}{F814W$^a$} & \multicolumn{1}{c}{$M_{V}^b$} &$(V-I)^b$&\multicolumn{1}{c}{$F_{X}/F_{O}$}\\
($\#$) & set~\# &\multicolumn{1}{c}{circle radius (\arcsec)} & \multicolumn{1}{c}{mag} & \multicolumn{1}{c}{(mag)} & \multicolumn{1}{c}{(mag)} &\multicolumn{1}{c}{(mag)} &\multicolumn{1}{c}{(mag)}\\
\hline
 1 & 3,5&0.15& ---&27.1$\pm$1.2&  26.9$\pm$1.9 &    -6.5$\pm$1.2   &         -0.1$\pm$2.2   &    186.0\\
2&2,4&0.10& --- &   25.5$\pm$0.7 & 24.0$\pm$0.4  &   -8.0$\pm$0.7   &      1.3$\pm$0.8     &      2.8\\
2&3,5&0.10& --- &   25.3$\pm$0.5 & 23.7$\pm$0.3  &   -8.2$\pm$0.5   &        1.4$\pm$0.6     &      2.3\\
3&1&0.10& 22.4$\pm$0.1  &    --- & --- &  ---    &     ---    &         ---\\
3&2,4&0.10& ---  &    23.3$\pm$0.2 & 22.1$\pm$0.1 &  -10.2$\pm$0.2    &        1.0$\pm$0.2      &       2.2\\
3&3,5&0.10&  ---&    22.2$\pm$0.1 & 21.2$\pm$0.1 &  -11.3$\pm$0.1    &        0.9$\pm$0.1      &       0.8\\
4&3,5&0.16&  ---&     27.4$\pm$1.2 & 26.9$\pm$1.5 &    -6.1$\pm$1.2    &             0.3$\pm$1.9   &      17.4\\
5&3,5&0.20&  ---&     23.5$\pm$0.2&  22.7$\pm$0.2 &  -9.9 $\pm$0.2  &              0.7$\pm$0.3     &       1.1\\
6&3,5& 0.20& ---   &  24.0$\pm$0.2 & 22.8$\pm$0.1 &  -9.5$\pm$0.2     &         1.0$\pm$0.3        &      1.1\\
\hline

\multicolumn{9}{l}{$^a$Observed magnitudes, not Galactic extinction corrected.} \\
\multicolumn{9}{l}{$^bM_{V}$ means Galactic extinction-corrected (N$_{H}$=6.98x10$^{20}$ cm$^{-2}$, N$_{H}$/E(B-V)=5.2x10$^{21}$ cm$^{-2}$ mag$^{-1}$ } \\
\multicolumn{9}{l}{and A/E(B-V)=3.315 for V and 1.940 for I) absolute magnitude at F555W band. Similarly, (V--I), } \\
\multicolumn{9}{l}{is the difference in extinction-corrected magnitudes in F555W, and F814W.}\\
\end{tabular}
} 
\end{center}

\vspace{15pt}

\begin{center}
\small{
\begin{tabular}{ccccccc}
\multicolumn{7}{c}{{\sc Table 4}} \\
\multicolumn{7}{c}{ X-Ray Spectral Parameters of the nucleus and the diffuse emission in NGC 5018} \\

\hline \hline

  \multicolumn{1}{c}{Source} & \multicolumn{1}{c}{Model} & \multicolumn{1}{c}{$N_H$} & \multicolumn{1}{c}{$\Gamma$} & \multicolumn{1}{c}{$kT_e$} & \multicolumn{1}{c}{$L_X^a$} & $\chi^2$/dof \\
 &  & \multicolumn{1}{c}{($10^{20}$ cm$^{-2}$)} & &\multicolumn{1}{c}{(keV)} & \multicolumn{1}{c}{($10^{39}$~erg~s$^{-1}$)}& \\

\hline

\vspace{5pt}

Diffuse &pha~$*$~Power-law&$65.9^{+17.6}_{-30.8}$&$6.8^{+1.0}_{-1.5}$&---&124.3$\pm$17.8$^b$&129.3/30\\
Diffuse &pha~$*$~Bremsstrahlung&$45.7^{+22.6}_{-26.4}$&---&$0.20^{+0.01}_{-0.01}$&43.7$\pm$27.8$^b$&103.9/30\\
Diffuse &pha~$*$~Mekal & $7.1^{+6.2}_{-6.5}$& ---&$0.44^{+0.05}_{-0.07}$& 7.6$\pm$1.7$^b$ & 51.2/30 \\
Diffuse &pha~$*$~(Mekal~+~Mekal)&$6.98^{+0.0}_{-0.0}$& $0.43^{+0.11}_{-0.13}$$^c$&$0.43^{+0.05}_{-0.05}$& 7.9$\pm$2.8$^b$ & 51.3/29 \\
 Diffuse &pha~$*$~(Mekal~+~Power-law) &6.98& $1.06^{+0.43}_{-0.45}$&$0.41^{+0.04}_{-0.04}$& 13.7$\pm$1.5$^b$ & 27.1/29 \\
\hline

Nucleus &pha~$*$~Mekal & $41.2^{+17.5}_{-23.5}$& ---&$0.26^{+0.07}_{-0.06}$& 20.7$\pm$3.9$^a$ & 33.5/8 \\
Nucleus &pha~$*$~Power-law & $59.6^{+20.8}_{-24.9}$ & $6.3^{+1.3}_{-2.8}$&---& 50.6$\pm6.9$$^a$ & 43.4/8\\
 Nucleus &pha~$*$~(Mekal~+~Power-law) & 6.98 & $1.56^{+0.46}_{-0.59}$&$0.39^{+0.08}_{-0.09}$&6.8$\pm$0.9$^a$ &3.8/7  \\

\hline 

\multicolumn{7}{l}{$^a$Intrinsic 0.5--8.0 keV luminosity in the 0-3$\arcsec$ region.}\\
\multicolumn{7}{l}{$^b$Intrinsic 0.5--8.0 keV luminosity in the 3$\arcsec$-20$\arcsec$ region.}\\
\multicolumn{7}{l}{$^c$Temperature of the second Mekal component in keV.}\\
\end{tabular}
} 
\end{center}

\vspace{15pt}

\begin{center}
\scriptsize{
\begin{tabular}{ccccccccccc}
\multicolumn{11}{c}{{\sc Table 5}} \\
\multicolumn{11}{c}{Radio continuum measurements of NGC~5018} \\

\hline \hline

\multicolumn{1}{c}{Source} & \multicolumn{1}{c}{R.A.} & \multicolumn{1}{c}{Dec.} &\multicolumn{1}{c}{S$^{a}_{P}$} & \multicolumn{1}{c}{S$^{a}_{T}$}&\multicolumn{1}{c}{3 cm}&\multicolumn{1}{c}{6 cm}&\multicolumn{1}{c}{13 cm}&\multicolumn{1}{c}{22 cm} & \multicolumn{1}{c}{Angular} & \multicolumn{1}{c}{Refe-} \\ 
 &\multicolumn{1}{c}{(1950)}&\multicolumn{1}{c}{(1950)}&\multicolumn{1}{c}{(mJy)}&\multicolumn{1}{c}{(mJy)}&\multicolumn{1}{c}{(mJy)}&\multicolumn{1}{c}{(mJy)}&\multicolumn{1}{c}{(mJy)}&\multicolumn{1}{c}{(mJy)}&\multicolumn{1}{c}{size~(\arcsec)}&\multicolumn{1}{c}{rence}\\
\hline
Nucleus & 13 10 19.9(1)&-19 15 14(10) & --- & 3.3$\pm0.50$ &---&---&---&---& L & 1\\
Nucleus & 13 10 20.00 & -19 15 11.0 & --- & 3.1$\pm0.17$&---&---&---&--- & $<$~20&2\\
Nucleus & 13 10 20.04 & -19 15 12.5 & 1.9$\pm0.1$  &1.9$\pm0.10$&---&---&---&--- & $<$~1&2 \\
Nucleus &13 10 20.05 & -19 15 11.9 &---&---&0.9$\pm0.1$&1.4$\pm0.1$&1.9$\pm0.1$&2.0$\pm0.1$& $<$~3&3\\
ULX~\#~6 & 13 10 25.3(0.7) & -19 15 26(10) & --- & 3.3$\pm0.50$&---&---&---&--- & L & 1\\
ULX~\#~6 & 13 10 25.50 & -19 15 26.0 & --- & 2.5$\pm0.17$&---&---&---&--- & $<$~20&2 \\
ULX~\#~6 & 13 10 25.45 & -19 15 21.5 & 4.0$\pm0.2$& 4.0$\pm0.20$&---&---&---&--- & $<$~1&2\\
ULX~\#~6&13 10 25.47&-19 15 21.2 &---&---&3.0$\pm0.1$&3.1$\pm0.1$&2.3$\pm0.2$&2.1$\pm0.3$& $<$~3&3\\
\hline

\multicolumn{11}{l}{$^a$20 cm, L: low resolution, 1:Kim \etal\ 1988, 2:Mollenhoff \etal\ 1992 and 3: this work}\\
\end{tabular}
} 
\end{center}


\begin{thebibliography}{}

\bibitem[]{1374}
Abramowicz, M. A., Czerny, B., Lasota, J. P., \& Szuszkiewicz, E. 1988, ApJ, 332, 646.
\bibitem[]{1374}
Begelman, M. C. 2002, ApJ, 568, L97.
\bibitem[]{1374}
Berendsen, S. G. H. 2000, MNRAS, 318, 599.
\bibitem[]{1374}
Bildsten, L., \& Deloye, C. J. 2004, ApJL, 607, 119.
\bibitem[]{1374}
Brandt, W. N. et al. 2001, AJ, 122, 2810.
\bibitem[]{1374}
Carollo, C. M. \& Danziger, I. J. 1994, MNRAS, 270, 743.
\bibitem[]{1374}
Colbert, E. J. M., \& Mushotzky, R. F. 1999, ApJ, 519, 89
\bibitem[]{1374}
Colbert, E. J. M., \& Ptak, A. F. 2002, ApJS, 143, 25
\bibitem[]{1374}
Colbert, E. J. M. et al. 2004, ApJ, 602, 231.
\bibitem[]{1374}
Fabbiano, G., \& White, N. 2003, astro-ph/0307077, to appear in
 ``Compact Stellar X-ray Sources,'' Cambridge University Press
 (eds., W. Lewin \& M. van der Klis)
\bibitem[]{1374}
Fabbiano, G. 2004, RMxAC, 20, 46.
\bibitem[]{1374}
Franco, J. \etal\ 1993, RevMAA, 27, 133
\bibitem[]{1025}
Georgakakis, A., Forbes, D. A. \& Norris, R. P. 2000, MNRAS, 318, 124.
\bibitem[]{1027}
Georganopoulos, M., Aharonian, F. A., \& Kirk, J. G., 2002 A\&A, 388, L25
\bibitem[]{1374}
Gilfanov, M. 2004, astro-ph/0403552, to appear in "Progress of Theoretical Physics", Proceedings of the workshop "Stellar-Mass, Intermediate-Mass, and Supermassive Black Holes", (Eds: K.Makishima and S.Mineshige.).
\bibitem[]{1374}
Goudfrooij, P., Hansen, L., Jorgensen, H. E., Norgaard,-Nielsen, H. U., de Jong, T., \& van den Hoek, L. B. 1994, A\&AS, 104, 179
\bibitem[]{1033}
Green, P. J. et al. 2004, ApJS, 150, 43.
\bibitem[]{1035}
Grimm, H. -J., Gilfanov, M., \& Sunyaev, R. 2003, MNRAS, 339, 793.
\bibitem[]{1374}
Guhathakurtha, P., et al. 1989, in ``The Interstellar Medium in External Galaxies'', ed. D. J. Hollenbach \& A. T. Harley, NASA Conf. Publication 3084, p. 26
\bibitem[]{1374}
Han, X.  \& Hjellming, R. M. 1992, ApJ, 400, 304.
\bibitem[]{1374}
Harris, W.E. 1996, AJ, 112, 1487
\bibitem[]{1374}
Hilker, M. \& Kissler-Patig, M. 1996, A\&A, 314, 357.
\bibitem[]{1374}
Homan, J. 2004, A\&A, 418, 255.
\bibitem[]{1374}
Humphrey, P. J. et al. 2003, MNRAS, 344, 134.
\bibitem[]{1374}
Hunter, D. A. et al. 2000, AJ, 120, 2383.
\bibitem[]{1374}
Irwin, J. A., Athey, A. E. \& Bregman, J. N. 2003, ApJ, 587, 356.
\bibitem[]{1374}
Irwin, J. A., Bregman, J. N.  \& Athey, A. E. 2004, ApJL, 601, L143
\bibitem[]{1374}
Kilgard, R. E. et al. 2002, ApJ, 573, 138.
\bibitem[]{1374}
Kim, D. -W.,  et al. 1988, ApJ, 330, 684.
\bibitem[]{1374}
King, A. R., Davies, M. B., ward, M. J., Fabbiano, G., \& Elvis, M. 2001, ApJ, 552, L109
\bibitem[]{1374}
King, A. R. 2002, MNRAS, 335, 13.
\bibitem[]{1374}
Lotz, J. M. 2004, ApJ, 613, 262.
\bibitem[]{1374}
Maccacaro, T. et al. 1988, ApJ, 326, 680.
\bibitem[]{1374}
Makishima, K., et al. 2000, ApJ, 535, 632
\bibitem[]{1374}
Malin, D. F. \& Carter, D. 1983, ApJ, 274, 534
\bibitem[]{1374}
Malin, D. F. \& Hadley, B. 1997, PASA, 14, 82
\bibitem[]{1374}
Miller, B. W. et al. 1997, AJ, 114, 2381.
\bibitem[]{1374}
Mollenhoff, C. \etal\ 1992, A\&A, 255, 35.
\bibitem[]{1374}
Mukai, K. 1993, Legacy, 3, 21.
\bibitem[]{1374}
Plewa, T. 1995, MNRAS, 275, 143
\bibitem[]{1374}
Prestwich, A. H., Irwin, J. A. \& Boroson, B. 2003, ApJ, 595, 719.
\bibitem[]{1374}
Roberts, T. P., \& Warwick, R. S. 2000, MNRAS, 315, 98.
\bibitem[]{1374}
Swartz, D. A., Ghosh, K. K., McCollough, M. L., Pannuti, T. G., Tennant, A. F., \& Wu, K. 2003, ApJS, 144, 213
\bibitem[]{1374}
Swartz, D. A., Ghosh, K. K., Tennant, A. F., \& Wu, K. 2004, ApJS, 154, 519.
\bibitem[]{1374}
Tennant, A. F. et al. 2005, in preparation
\bibitem[]{1374}
Terashima, Y. \& Wilson, A. S. 2003, ApJ, 583, 145
\bibitem[]{1374}
Thronson, H. A. \& Bally, J. 1987, ApJL, 319, L63.
\bibitem[]{1374}
van der Marel, R. P. 2003, astro-ph/0302101
\bibitem[]{1374}
Veron-Cetty M.P. \& Veron P.2003 A\&A, 412, 399.
\bibitem[]{1374}
Whitmore, B. C. et al. 1999, AJ, 118, 1551.
\bibitem[]{1374}
Worrall, D. M. \& Birkinshaw, M. 2004, astro-ph/0410297.
\bibitem[]{1374}
Zaggia, S. et al. 1999, A\&AS, 137, 75.
\end{thebibliography}
\end{document}